\documentclass[onecolumn, UTF8]{article}
\usepackage{arxiv}

\usepackage{graphicx}
\usepackage{subfigure}
\usepackage{tabularx}
\usepackage{multirow}
\usepackage{enumerate}
\usepackage{CJKutf8}
\usepackage[colorlinks,linkcolor=red,anchorcolor=blue,citecolor=green,CJKbookmarks=True]{hyperref}
\usepackage[linesnumbered,ruled,lined]{algorithm2e}
\usepackage{cite}
\begin{document}

\title{A Multitask Deep Learning Approach for User Depression Detection on Sina Weibo}

\author{ \href{https://orcid.org/0000-0002-2735-6977}{\includegraphics[scale=0.06]{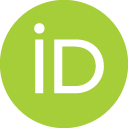}\hspace{1mm}Yiding Wang}\\
	College of Cybersecurity\\
	Sichuan University\\
	Chengdu, China 610207 \\
	\texttt{wangyiding@stu.scu.edu.cn} \\
	\And
	Zhenyi Wang \\
	College of Cybersecurity\\
	Sichuan University\\
	Chengdu, China 610207 \\
	\texttt{wangzhenyi@stu.scu.edu.cn}
\And
	Chenghao Li \\
	College of Cybersecurity\\
	Sichuan University\\
	Chengdu, China 610207 \\
	\texttt{2017141531008@stu.scu.edu.cn}
    \And
    Yilin Zhang \\
	College of Cybersecurity\\
	Sichuan University\\
	Chengdu, China 610207 \\
	\texttt{2017141482187@stu.scu.edu.cn}
    \And
    \href{https://orcid.org/0000-0003-1197-5906}
    {\includegraphics[scale=0.06]{orcid.png}\hspace{1mm}Haizhou Wang}\thanks{Corresponding author: H. Wang (whzh.nc@scu.edu.cn)} \\
	College of Cybersecurity\\
	Sichuan University\\
	Chengdu, China 610207 \\
	\texttt{whzh.nc@scu.edu.cn}
}



\hypersetup{
pdftitle={A Multitask Deep Learning Approach for User Depression Detection on Sina Weibo},
pdfsubject={cs.AI},
pdfauthor={Yiding Wang, Zhenyi Wang, Chenghao Li, Yilin Zhang, Haizhou Wang},
pdfkeywords={Depression detection, Online social network, Feature engineering, Deep neural network, Multitask learning.},
}
\maketitle

\begin{abstract}
In recent years, due to the mental burden of depression, the number of people who endanger their lives has been increasing rapidly. The online social network (OSN) provides researchers with another perspective for detecting individuals suffering from depression. However, existing studies of depression detection based on machine learning still leave relatively low classification performance, suggesting that there is significant improvement potential for improvement in their feature engineering. In this paper, we manually build a large dataset on Sina Weibo (a leading OSN with the largest number of active users in the Chinese community), namely Weibo User Depression Detection Dataset (WU3D). It includes more than 20,000 normal users and more than 10,000 depressed users, both of which are manually labeled and rechecked by professionals. By analyzing the user's text, social behavior, and posted pictures, ten statistical features are concluded and proposed. In the meantime, text-based word features are extracted using the popular pretrained model XLNet. Moreover, a novel deep neural network classification model, i.e. FusionNet (FN), is proposed and simultaneously trained with the above-extracted features, which are seen as multiple classification tasks. The experimental results show that FusionNet achieves the highest F1-Score of 0.9772 on the test dataset. Compared to existing studies, our proposed method has better classification performance and robustness for unbalanced training samples. Our work also provides a new way to detect depression on other OSN platforms.
\end{abstract}

\keywords{Depression detection \and Online social network \and Feature engineering \and Deep neural network \and Multitask learning}

\section{Introduction}
\subsection{Background}
With the rapid development of the online social network (OSN) such as Twitter and Facebook, people are more frequently using the OSN to express opinions and emotions. It provides researchers with a novel and effective way to detect the mood, communication, activity, and social behavior pattern of individuals \cite{de2013predicting}. In the past decade, researchers in various fields have conducted quantitative analyses of different illnesses and mental disorders based on the OSN platform \cite{wang2017detecting,karmen2015screening,saravia2016midas,zhang2018optimizing,lin2014user,balani2015detecting,cheng2017assessing}. Sina Weibo (hereinafter referred to as ``Weibo'') is the most popular OSN in the Chinese community \cite{gao2012comparative}. A statistic shows the number of Weibo's monthly active users have reached more than 480 million in the second quarter of 2019\footnote{https://www.statista.com/statistics/941456/china-number-of-sina-weibo-users/}.

Major depressive disorder, referred to as depression, is a common mental disease. According to a survey of the World Health Organization (WHO)\footnote{https://www.who.int/en/news-room/fact-sheets/detail/depression}, more than 300 million people worldwide suffer from depression. Depression can cause great psychological pain, even suicidal tendencies. Moreover, evidence from a health action plan of WHO\footnote{https://www.who.int/mental\_health/action\_plan\_2013/en/} shows that people suffering from depression are much more likely to end their life prematurely than the general population. Despite the current availability of psychotherapy, medical therapy, and other modalities for the treatment of depression, 76\%-85\% of patients in low- and middle-income countries remain untreated. The emergence of this phenomenon is not only the lack of medical resources but also the inability to make an accurate assessment in the early stage of depression, which leads to a large number of people with depression difficult to get diagnosis and treatment timely \cite{shen2017depression}.

Pictures, text, videos, and other information posted on the OSN can reflect feelings of worthlessness, guilt, helplessness, and self-hatred, which can help researchers to specifically analyze and characterize depressed individuals \cite{de2013predicting,shen2017depression,shen2018cross,gui2019cooperative}. However, there are some insurmountable problems in online depression detection using traditional analyzing methods. They often focus on analyzing the characteristics of users with depression rather than constructing predictive models. Therefore, it is difficult to give timely prediction results of new depressed users. Moreover, they are incapable to deal with a large number of instant interactive user data.

With the rapid development of artificial intelligence technologies, machine learning approaches have made great contributions to the detection of depression \cite{suhara2017deepmood,ma2016depaudionet,yang2017multimodal,dibekliouglu2017dynamic,zhu2017automated}.
An automated depression detection model based on machine learning usually needs to analyze various information such as tweets, pictures, videos, social activity data of users. Then, it gives the classification results of the predicted objects, most of which are presented as a binary result of normal or depressive. If an individual is predicted for a potential depressive tendency, further resources and assistance can be provided, including later medical and psychological diagnoses. Such heuristic learning approaches are quite effective for helping in the early detection of depression \cite{guntuku2017detecting} since they are capable of handling a large number of instant interactive user data.

\subsection{Challenges}
However, current approaches to online depression detection still face many unresolved challenges.

Firstly, many current studies are not user-oriented modeling \cite{cohan2018smhd,deshpande2017depression,al2019depression}. Those works usually aim to analyze and model the language style of the user. Through sentiment analysis and feature engineering of the tweet text, a classification model is developed to detect whether a specific tweet has a depressive tendency. These works analyzed fine-grained features and achieved pretty good results. However, such results cannot be directly applied to user-level depression detection, or it may lead to an incorrect prediction.

Second, in several existing studies \cite{de2013predicting,cohan2018smhd,trotzek2018utilizing,mustafa2020multiclass,stankevich2018feature}, the size of the dataset used for modeling is insufficient, with only a few hundred to a few thousand data samples being used. Because of the difficulty of accurately obtaining and labeling depressed samples, researchers usually choose to construct small datasets or directly cited datasets from other works. As a consequence, the trained model fails to reach good generalization performance, thus hard to accurately predict depressed users on the OSN.

Moreover, not enough studies of user depression detection have been proposed on Weibo compare to Twitter and Facebook. To the best of our knowledge, there is no published large Weibo user depression detection dataset available currently.

Finally, many of the existing proposed models still do not reach a high level of classification performance, i.e. an F1-Score of 90\% and above. Thus, these models need to be further improved to achieve better performance.

\subsection{Contributions}
Given the above problems and challenges, we hereby summarize the contributions of our work as below:
\begin{itemize}
  \item \textbf{We build and publish a large-scale labeled dataset - Weibo User Depression Detection Dataset (WU3D)\footnote{https://github.com/aidenwang9867/Weibo-User-Drpession-Detection-Dataset}}. WU3D includes more than 10,000 depressed users and more than 20,000 normal users, each of which contains enriched information fields, including tweets, the posting time, posted pictures, the user gender, etc. This dataset is labeled and further reviewed by professionals.
  \item \textbf{We summarize ten features of depressed users, four of which are the first to be proposed.} Different from some existing work that directly using the information fields as features, we made statistical analyzes of all the proposed features. These features show significant distribution differences between depressed and normal users in our experiments.
  \item \textbf{We construct a Deep Neural Network (DNN) classification model, i.e. FusionNet.} It implements a multitask learning strategy to process text-based word vectors and statistical features simultaneously. Experimental results show that it achieves both the highest classification performance and the best robustness to unbalanced training samples.
\end{itemize}

The subsequent sections of this paper are organized as follows. In Section \uppercase\expandafter{\romannumeral2}, related work and achievements in the field of depression detection on OSNs are introduced and analyzed. The proposed framework is elaborated in Section \uppercase\expandafter{\romannumeral3}. Furthermore, Section \uppercase\expandafter{\romannumeral4} gives the significance evaluation of statistical features and the performance comparison experiments of several classification models (including our proposed FusionNet). At the last of the paper, Section \uppercase\expandafter{\romannumeral5} summarizes our work and discusses directions for future work.

\section{Related Work}
The current methods for online depression detection mainly include two directions. \textbf{(i)} Manually extracting features and building Traditional Machine Learning (TML) models for classification. \textbf{(ii)} Using Deep Learning (DL) approaches to automatically extract features and constructing deep neural network models as classifiers.

Among them, some of the research that uses DL also introduces TML methods to further improve their model performance. The research of each approach will be introduced below respectively.

\subsection{Detection Approaches based on Traditional Machine Learning}
Mining depression users based on TML mostly uses features, i.e. numeric vectors that have been manually analyzed and extracted from users to represent the predicted object (a user, a tweet, a posted picture, etc.) \cite{guntuku2017detecting}.

Choudhury et al. \cite{de2013predicting} presented a pioneering work in this field of research. They explored potential user behavior to perform a user-oriented depression detection. By measuring behavioral attributes on Twitter users relating to social engagement, emotion, language, and linguistic styles, they discovered useful signals for characterizing depression. Although their trained classifiers did not achieve high classification performance, as a pioneering work in this field, they provided a detailed feature engineering analysis process and a clear modeling approach.

Wang et al. \cite{wang2013depression} undertook further research using data from Twitter and Weibo. Compared with the work of \cite{de2013predicting} that made a more comprehensive feature analysis, this study implemented a sentiment analysis approach and proposed man-made rules by utilizing vocabulary to measure depressive tendencies of tweets. Their work indicated that text-based features play a crucial role in online depression detection.

Deshpande et al. \cite{deshpande2017depression} proposed a representation learning method based on natural language processing (NLP) to model the text information on Twitter. Different from the previously mentioned work \cite{de2013predicting,wang2013depression}, they used the Bag of Words (BOW) algorithm to represent the tweet text as a sparse vector, allowing the classifier to automatically learn latent features. Their trained Naive Bayes (NB) classifier reached an F1-Score of 0.8329, while the Support Vector Machine (SVM) classifier only reached an F1-Score of 0.7973.

After, Shen et al. \cite{shen2017depression} proposed an advanced detecting approach that can be used to detect depressed users timely. They constructed a well-labeled depression detection dataset on Twitter, which had been widely used by subsequent researchers. In the meantime, they extracted six depression-related feature groups covering the text, social behavior, and posted pictures. Their proposed multimodal depressive dictionary learning (MDL) approach can effectively learn the latent and sparse representation of user features. Experiments showed their proposed MDL model achieved an F1 of 0.85, indicating that the dictionary learning strategy and the ensemble of multimodal is quite effective.

In recent years, more TML-based work has begun to emerge \cite{al2019depression,mustafa2020multiclass,stankevich2018feature}. In particular, Mustafa et al. \cite{mustafa2020multiclass} implemented Frequency-Inverse Document Frequency (TF-IDF) algorithm to weight the words in tweets. Their trained classifier based on an one-dimensional convolution neural network (CNN-1D) achieved an F1-Score of 0.89. Their work is the first to introduce a neural network model for detecting depressed users on the OSN.

\subsection{Detecting Approaches based on Deep Learning}
Modeling approaches based on DL are mainly for jointly considering user social behaviors and multimedia information such as the text, pictures, videos, etc. Among them, the modeling of the text information is the main research direction. Researchers have adopted NLP approaches to embed text into a high dimensional continuous vector to automatically mine word features. Some work has also fused manually extracted features into DNN classifiers as part of the input, or integrated traditional classifiers with DNN classifiers to improve performance. These multimodal and ensemble approaches have proven to be an effective way to accomplish various tasks on social network analysis including depression detection \cite{huang2019multimodal}.

Several DNN classifiers that have achieved significant performance in the NLP classification task were selected and evaluated by Orabi et al. \cite{orabi2018deep}. They used a pretrained Word2Vec \cite{mikolov2013distributed} model to embed the text of tweets. Their experimental results showed that the CNN-1D with a max-pooling structure reported the highest performance. Compared to other recurrent structures including the recurrent neural network (RNN), the Long Short-Term Memory (LSTM) neural network \cite{chorowski2014end,hochreiter1997long}, CNN-based models performed better in the task of depression detection.

Then, Sadeque et al. \cite{sadeque2018measuring} proposed a latency-weighted F1 metric and applied it in a novel sequential classifier based on the Gated Recurrent Units (GRU). They treated all the text of tweets as documents and input them to the classifier asynchronously, named ``post-by-post'' strategy. It allows the model to decide the depressive tendency of a user after each tweet is scanned. Thus, it somehow avoids the time consumption of scanning too many tweets under a certain and obvious depressed user (e.g, a user with 200 tweets recording its anti-depressant experience). This approach can scan and detect depressive tendencies of tweets more efficient.

Later, based on the prior work \cite{shen2017depression}, Shen et al. \cite{shen2018cross} discovered that the current research on a specific OSN may be unsuitable and not universal for depression detection on other platforms. Thus, they proposed a cross-domain DNN model with Feature Adaptive Transformation \& Combination (DNN-FATC) strategy that can consider features of several aspects comprehensively and transfer the relevant information across heterogeneous domains.

Recently, more studies based on DL have been widely proposed. Gui et al. \cite{gui2019cooperative} further discussed the change of classification accuracy of the model under the different proportions of depressed users and pointed out that the highest accuracy can be achieved when the proportion of normal and depressed user samples is close to balance. Moreover, they implemented a reinforcement learning (RL) approach to further improve the performance of the model. Lin et al. \cite{lin2020sensemood} used a popular pretrained model, i.e. BERT \cite{devlin2018bert}, to embed word vectors. Its hidden layer output was extracted to fuse both text and image features to further accomplish the downstream classification task.

\section{Framework and Methodology}
\begin{figure*}[htbp]   
    \centering
    \includegraphics[scale=0.27]{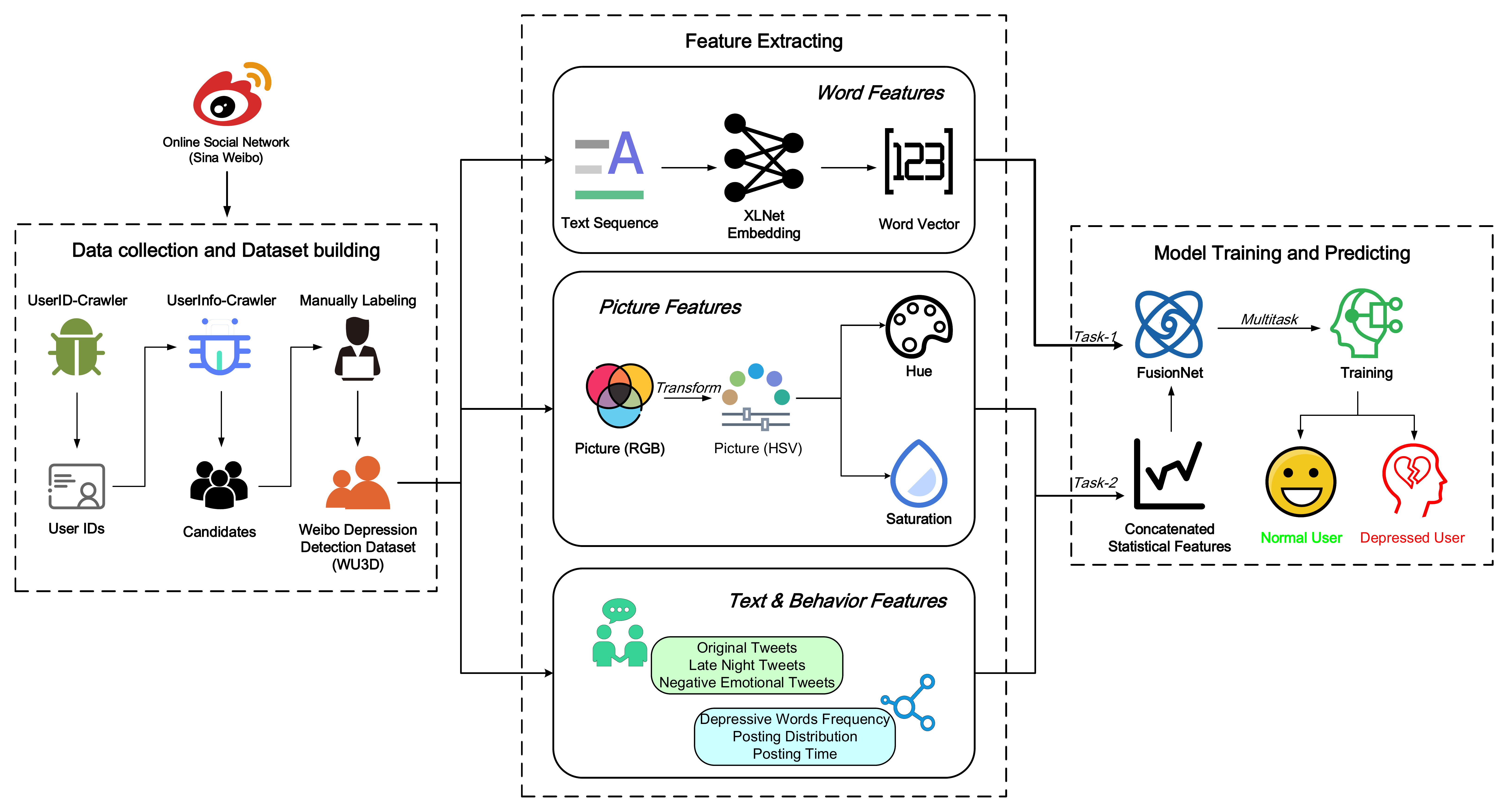}
    \caption{The Framework of the Proposed Method}
    \label{framework}
\end{figure*}

To detect the depressed users on the OSN more effectively, we propose a novel framework, as shown in Fig. \ref{framework}. This framework mainly consists of three parts.

\begin{enumerate}[i.]
  \item \emph{User data collection and labeling.} This module contains two independent crawler systems (\emph{UserID-Crawler} and \emph{UserInfo-Crawler}), which are used to collect user samples on Weibo. Then, it is responsible for filtering and labeling the collected data to construct the Weibo User Depression Detection Dataset (WU3D).
  \item \emph{Feature extracting.} This module is in charge of extracting the user's text information including nicknames, profiles, tweet text, and concatenating them into a long text sequence. Then, the sequence is input to the XLNet \cite{yang2019xlnet} pretrained model to obtain embedded word features. In the meantime, this module extracts statistical features of user text, social behavior, posted pictures. Finally, these features are jointly input into the classification model.
  \item \emph{Model Training and predicting.} The module implements a depression detection model based on DNN, namely FusionNet, which receive features input from the \emph{Feature Extracting} module. The proposed FusionNet can be trained in a multitask learning mode, in which word vectors and statistical features can be used jointly to optimize the classifier in each training step.
\end{enumerate}

The following parts of this section will elaborate on the theoretical construction and implementation methods of these modules, respectively.

\subsection{User Data Collection and Labeling}
\subsubsection{Data collection}
A user ID can be used to uniquely identify a user. With a user ID, the crawler can access the user's home page and collect information from it. First of all, \emph{UserID-Crawler} is constructed to collect user IDs of depressed candidates. The API provided by Weibo official\footnote{https://open.weibo.com/wiki/API} is used to obtain as accurate information as possible. Our strategies for collecting user IDs of depressed candidates include:

(i) Collecting data from the Weibo Super Topic of ``\begin{CJK*}{UTF8}{gbsn}抑郁症\end{CJK*}'' (``Depression'' in English). The Super Topic is a social group on Sina Weibo that gathers users with common interests. It has been proved that individuals who share the same background are more likely to trust each other, thus will gather to form aggregations \cite{meo2015trust}. According to our investigation and analysis, there are a large number of active depressed users posting under the topic of ``Depression''. Collecting data in this way can greatly improve the efficiency of gathering depressed user samples. Therefore, \emph{UserID-Crawler} collects depressed candidates under this topic and forms a list of their user IDs.

(ii) Collecting data through the function of ``\begin{CJK*}{UTF8}{gbsn}微博搜索\end{CJK*}'' (``Weibo Search'' in English) provided by Weibo official\footnote{https://s.weibo.com/}. We use high-frequency words including ``\begin{CJK*}{UTF8}{gbsn}抑郁症\end{CJK*}'' (``Depression'' in English), ``\begin{CJK*}{UTF8}{gbsn}自杀\end{CJK*}'' (``Suicide'' in English), ``\begin{CJK*}{UTF8}{gbsn}痛苦\end{CJK*}'' (``Pain'' in English) and the late night time period (from 0:00 a.m. to 6:00 a.m.) as two main search conditions to crawl user IDs for collecting more depressed candidates.

\begin{figure}[htbp]
    \centering
    \includegraphics[scale=0.7]{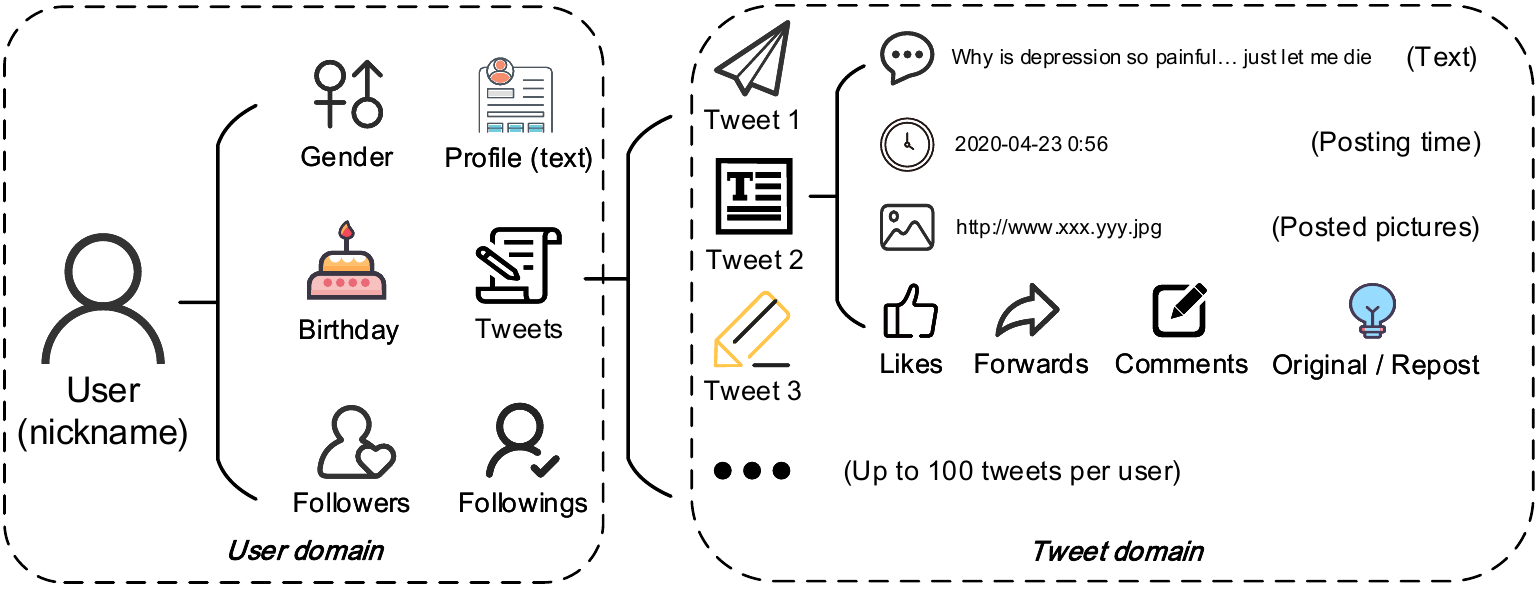}
    \caption{The Data Structure of Candidates and WU3D (per user)}
    \label{datafield}
\end{figure}

Through the above two crawling strategies, we have collected sufficient user IDs of depressed candidates. Then, with the user ID list, \emph{UserInfo-Crawler} is implemented to collect detailed user information from its personal homepage. The specific information fields collected by \emph{UserInfo-Crawler} are shown in Fig. \ref{datafield}.

We divide the information for each user sample into two domains: the user domain and the tweet domain. The user domain contains the user's gender, birthday, profile (a short text of the user's self-description), the number of followers, the number of followings, and the list of tweets. For each tweet in the tweet domain, it contains the tweet text, the posting time, posted pictures, the number of likes, the number of forwards, the number of comments, and an identifier that identifies whether the tweet is original or not.

For normal candidates, we use \emph{UserID-Crawler} to collect them under four Super Topics including ``\begin{CJK*}{UTF8}{gbsn}日常\end{CJK*}'' (``Daily'' in English), ``\begin{CJK*}{UTF8}{gbsn}正能量\end{CJK*}'' (``Positive Energy'' in English), ``\begin{CJK*}{UTF8}{gbsn}榜姐每日话题\end{CJK*}'' (``Daily Topic'' in English), ``\begin{CJK*}{UTF8}{gbsn}互动\end{CJK*}'' (``Interaction'' in English) to form a list of normal candidate IDs. Then, the more detailed user information is collected through \emph{UserInfo-Crawler} to form the same data fields and structure as the depressed candidates. Based on the previous steps, we have collected 125,479 depressed candidates and 65,913 normal candidates.

\subsubsection{Data filtering and labeling}
Automated scripts are implemented to filter out non-personal accounts by identifying the user's \emph{``account type''} field, including marketing accounts, official accounts, and social bots.

The automated filtered normal candidates is labeled as normal users directly without further manual labeling. For depressed users, we invite professional data labelers to complete the data labeling process of depressed candidates. To ensure that the results are highly reliable, the labeled data has been reviewed twice by psychologists and psychiatrists. The principles of the data labeling can be described as follows:

\begin{enumerate}[i.]
  \item Depressed candidates with a self-reported history of depression, confirmed diagnosis, currently taking antidepressants, and recording antidepressant experiences in multiple tweets will be labeled as depressed users.
  \item If a candidate's tweets have repeatedly appeared the content of describing psychological suffering, mental anguish, and strong suicide intention, the user will be identified as depressed.
  \item If the posted pictures of a candidate repeatedly involve or show bloodshed and self-harming content and the tweet text includes keywords such as ``\begin{CJK*}{UTF8}{gbsn}抑郁\end{CJK*}'' (``Depression'' in English) and ``\begin{CJK*}{UTF8}{gbsn}自残\end{CJK*}'' (``Self-harming'' in English), the candidate will be identified as a depressed one.
  \item Candidates who partially meet the above conditions but have too many unrelated contents such as forwarding lottery prizes, receiving red envelopes, advertising information, will be directly discarded.
\end{enumerate}

Therefore, the target dataset, i.e. WU3D, is constructed. It contains both labeled normal and depressed users. The specific information of the candidates and WU3D are given in Table \ref{datasetstatistics}. We counted the normal sample, the depressed sample, and the total for each of the two types. In particular, we give a detailed number of users and their posted tweets, and posted pictures.

\begin{table}[htbp]
    \centering
    \caption{Dataset statistics}
    \begin{tabular}{clccc}
        \hline
        \multicolumn{1}{c}{\textbf{Dataset}} &
        \multicolumn{1}{c}{\textbf{Category}} & \multicolumn{1}{c}{\textbf{User}} & \multicolumn{1}{c}{\textbf{Tweet}} & \multicolumn{1}{c}{\textbf{Picture}} \\ \hline
        \multicolumn{1}{c}{}          & Depressed             & 125,479           & 5,478,806  & 2,354,701 \\
        \multicolumn{1}{c}{Candidates} & Normal                & 65,913            & 4,927,904  & 3,631,537 \\
        \multicolumn{1}{c}{}          & Total                 & 191,392           & 10,406,710 & 5,986,238 \\ \hline
        \multicolumn{1}{c}{}          & Depressed             & 10,325            & 408,797    & 160,481  \\
        \multicolumn{1}{c}{WU3D}      & Normal                & 22,245            & 1,783,113  & 1,087,556  \\
        \multicolumn{1}{c}{}          & Total                 & 32,570            & 2,191,910  & 1,248,037 \\ \hline
    \end{tabular}
    \label{datasetstatistics}
\end{table}

All of the candidates were collected from March 2020 to May 2020. A total of over 200,000 user samples were collected, including 125,479 depressed candidates and 65,913 normal candidates. After strict data filtering and labeling, the number of depressed users in WU3D reached 10,325, with the retention rate of 8.23\%; the number of normal users reached 20,338, with the retention rate of 29.34\%. The total user data retention rate was 15.50\%.

\subsection{Statistical Feature Engineering}
Several previous studies have defined features that are quite effective for detecting depressed users, such as the proportion of late-night tweets, the proportion of original tweets, and the mean value of hue and saturation. Based on their work, we firstly perform feature engineering of user features in three aspects: the user text, social behavior, and posted pictures. We then summarize ten user-level features including four brand-new proposed and two modified. These features are extracted using statistical approaches, including the scale, the mean value, the standard deviation, etc. In Table \ref{symboldef}, symbol definitions that appear in this section and subsequent sections are given.

\begin{table}[htbp]
    \centering
    \caption{Variable and Function Symbol Definitions}
    \begin{tabular}{|c|p{9.8cm}|}
\hline
\textbf{Symbol} & \textbf{Description}                                                 \\ \hline
$P$               & The posted tweet set of a user, including original and repost tweets.                                             \\ \hline
$t_p$               & The posting time of a tweet.                                         \\ \hline
$l_e$               & The emotional label of a tweet.      \\ \hline
$n_d$               & The number of depression-related words in a tweet.                   \\ \hline
$T$ & A set of all user text information, including the nickname, the profile, and the tweet text.                                              \\ \hline
$\pi$               & The posted pictures set of a user.                                   \\ \hline
$\mu={(h_\mu,s_\mu,v_\mu)}$ & The dominant color of a picture, a ternery contains hue, saturation and brightness of HSV color space. One picture has one dominant color. \\ \hline
$\overline{X}$               & The mean sample value of an attribute.                               \\ \hline
$S_\xi$                & The concatenated user long text sequence.                    \\ \hline
$\Delta$               & The max length of the long text sequence $S_\xi$.                            \\ \hline
$C$               & The function that calculates the number of elements in a set. \\ \hline
$L$      & The loss function of a neural network.                               \\ \hline
$\Theta$               & The parameter set of a neural network.                               \\ \hline
$y$               & The true label of a user data sample.                                     \\ \hline
$\hat{f}$               & The objective function of a neural network. It inputs a user's feature vector and outputs the predicted label.               \\ \hline
$J$               & The joint optimization function of a neural network.                 \\ \hline
\end{tabular}
    \label{symboldef}
\end{table}

Descriptions of these features are shown in Table \ref{featuredefine}. The features are divided into three groups, including text-based features, social behavior-based features, and picture-based features. Here, we give specific descriptions and formulas to calculate each feature.

\begin{table*}[htbp]
    \centering
    \caption{Manually extracted user features}
    \begin{tabular}{cp{4.5cm}lc}
        \hline
        \textbf{Group} & \multicolumn{1}{c}{\textbf{Feature name}} & \multicolumn{1}{c}{\textbf{Symbol}} & \multicolumn{1}{c}{\textbf{Source}} \\ \hline
        \multirow{2}{*}{Text: $\Psi$}             & Proportion of negative emotional tweets & $\psi_{NP}$ & First proposed in our work                \\
                                                  & Frequency of depression-related words   & $\psi_{FDW}$ & \cite{de2013predicting,wang2013depression,stankevich2018feature,shen2018cross,shen2017depression,lin2014user,cheng2017assessing}, modified in our work \\ \hline
        \multirow{4}{*}{Social  behavior: $\Phi$} & Proportion of original tweets           & $\phi_{POP}$ & \cite{wang2013depression,shen2018cross,lin2014user}            \\
                                                  & Proportion of late-night posting        & $\phi_{PLNP}$ & \cite{wang2013depression,stankevich2018feature,shen2017depression,shen2018cross,lin2014user}, modified in our work       \\
                                                  & Posting frequency (per week)            & $\phi_{PF}$ & First proposed in our work                \\
                                                  & Standard deviation of posting time      & $\phi_{SDPT}$ & First proposed in our work                \\ \hline
        \multirow{4}{*}{Picture: $\Gamma$}          & Frequency of picture posting            & $\gamma_{FPP}$ & First proposed in our work                \\
                                                  & Proportion of cold color-styled pictures  & $\gamma_{PCP}$ & \cite{shen2018cross}                \\
                                                  & Standard deviation of hue               & $\gamma_{SDH}$ & \cite{shen2017depression,shen2018cross,lin2014user}, modified in our work            \\
                                                  & Standard deviation of saturation        & $\gamma_{SDS}$ & \cite{shen2017depression,shen2018cross,lin2014user}, modified in our work            \\ \hline
    \end{tabular}
    \label{featuredefine}
\end{table*}

\subsubsection{Text-based features}
\textbf{Proportion of negative emotional tweets}. In previous works for Twitter \cite{de2013predicting,sadeque2018measuring}, by considering the number of tweets with negative emotions, they have achieved good results in distinguishing depressed users from normal ones. Rather than directly using its ``number'', we use the ``proportion'' calculation method to normalize the feature. Although depressive tendencies do not fully equate to the expression of negative emotions, when the proportion of tweets with negative emotions reaches a certain level, it can reflect that the user's mental state is depressed and painful, thus can reveal a tendency of depression. We use the Text Sentiment Analysis API of the Baidu Smart Cloud Platform\footnote{http://ai.baidu.com/tech/nlp/sentiment\_classify} to label all the original tweets. The API returns three emotional labels: 0 for negative, 1 for neutral, and 2 for positive. We retain the negative emotions of label 0 and summarize label 1 and 2 as a category of non-negative emotions. For all the original tweets under each user, we give the defination of $\psi_{NP}$ in equation (\ref{psi_np}), in which $C_{(P_o)}$ is the total number of original tweets, $C_{(l_e)}$ is the total number of original tweets with negative emotions:
\begin{equation}\label{psi_np}
    \psi_{NP}=\frac{1}{C_{(P_o)}} \times C_{(l_e)}, \ \psi_{NP}\in[0, 1]
\end{equation}

\textbf{Frequency of depression-related words}. Researchers have focused on the lexical and semantic analysis of the tweet text and quantified these features by self-constructing or quoting depression-related semantic lists \cite{de2013predicting,wang2013depression,stankevich2018feature,shen2017depression,shen2018cross,lin2014user,cheng2017assessing}. The results of the existing studies indicate that features based on high-frequency depression keywords can significantly improve the classification performance. We use ``frequency'' to describe how frequently depression-related words appeared in a user's tweets, reflecting its potential depressive tendencies. In our previous investigation and analysis on Weibo, we summarized a list of high-frequency words for depression. Here, it is used to calculate the frequency of depressive words in users' original tweets. The number of occurrences of depression-related words of each tweet $n_{d}$ is counted by matching the keyword list. Then, $\psi_{FDW}$ is calculated by:
\begin{equation}\label{psi_fdw}
    \psi_{FDW}=\frac{1}{C_{(P_o)}} \times \sum_{i=1}^{C_{(P_o)}}n_{d_i}, \ \psi_{FDW}\in[0, \infty)
\end{equation}

\subsubsection{Social behavior-based features}
\textbf{Proportion of original tweets}. Several related works have proved that depressed users are more likely to post a large number of original tweets to express their negative psychological state, with relatively few repost tweets \cite{wang2013depression,shen2018cross,lin2014user}. Therefore, we use the proportion of original tweets to distinguish between depressed users and normal users. Here we use $C_{(P)}$ to calculate the total number of tweets, including orignal tweets and repost tweets. Then, $\phi_{POP}$ is defined by:
\begin{equation}\label{psi_fdw}
    \phi_{POP}=\frac{1}{C_{(P)}} \times C_{(P_o)}, \ \phi_{POP}\in[0, 1]
\end{equation}

\textbf{Proportion of late-night posting}. Late night is a time when depressive symptoms more frequently attack, thus depressed users tend to be more likely to post tweets in this period \cite{wang2013depression,stankevich2018feature,shen2017depression,shen2018cross,lin2014user}. Moreover, the late-night period is the time when normal users sleep and rest. They rarely use social tools during this time and therefore send very few tweets. We use the proposed feature ``Tweet Time'' in Ref. \cite{shen2018cross} and make minor modifications. The time range of 0:00-6:00 is adopted as the late-night period. Moreover, all the tweets of a user are used to calculate, including original and repost ones. Then, $\phi_{PLNP}$ is given by equation (\ref{phi_plnp}), in which $C_{(t_p)}$ is used to calculate the total number of tweets posted in the late night time period from 0:00 a.m. to 6:00 a.m.
\begin{equation}\label{phi_plnp}
    \phi_{PLNP}=\frac{1}{C_{(P_o)}} \times C_{(t_p)}, \phi_{PLNP} \in [0, 1]
\end{equation}

\textbf{Posting frequency (per week)}. The previous study for Twitter \cite{wang2013depression} has found that there is also a difference in posting frequency between normal and depressed users. Depressed users tend to post large numbers of tweets when they are suffering from depression and heavily rely on social media to express their painful feelings. Moreover, ``Week'' is a moderate time size and has stronger periodicity than ``Month''. We take the earliest posting time and latest posting time as an interval, count the total number of tweets $C_{{(P}_{int})}$ during this interval, and then divide it by 7 to get the weekly frequency value. Thus, $\phi_{PF}$ can be represented by equation (\ref{phi_pf}):
\begin{equation}\label{phi_pf}
    \phi_{PF} = \frac{1}{7} \times C_{{(P}_{int})}
\end{equation}

\textbf{Standard deviation of posting time}. The posting time of depressed users tends to be concentrated in the late-night, while the relative distribution of post time of normal users is more discrete within a day \cite{wang2013depression,shen2017depression,shen2018cross,lin2014user}. Hence, we use the standard deviation to describe this phenomenon, in order to reflect the aggregation trend of users' posting time. The smaller the value of this feature, the more likely that user would post at a specific time period. Here, we consider all the original and repost tweets. The mean value of posting time $\overline{X}_{SDPT}$ is calculated by:
\begin{equation}\label{x_sdpt}
  \overline{X}_{SDPT}=\frac{1}{C_{(P)}} \times \sum_{i=1}^{C_{(P)}}t_{P_i}
\end{equation}
Then, $\phi_{SDPT}$ can be defined as:
\begin{equation}\label{phi_sdpt}
    \phi_{SDPT}=\sqrt{\frac{1}{C_{(P)}} \times \sum_{i=1}^{C_{(P)}}(t_{P_i}-\overline{X}_{SDPT})^{2}}
\end{equation}

\subsubsection{Picture-based features}
\textbf{Frequency of picture posting}. In existing works for Twitter and Weibo \cite{shen2018cross,lin2014user}, ``Tweet with pictures'' is categorized into ``Tweet type'' to measure how often users post pictures in their tweets and has achieved good performance. Based on our prior research on Weibo, we also found that depressed users were more likely to use a lot of text to express their feelings and mental states, thus have fewer posted pictures than normal users. Therefore, we propose this feature to reflect users' habit of posting pictures. $C_{(\pi)}$ represents the total number of posted pictures. Then, we calculate $\gamma_{FPP}$ by:
\begin{equation}\label{gamma_fpp}
  \gamma_{FPP}=\frac{1}{C_{(P_o)}} \times C_{(\pi)}
\end{equation}

\textbf{Proportion of cold color-styled pictures}. Studies for Twitter \cite{shen2018cross,lin2014user} and Weibo \cite{shen2017depression,shen2018cross,lin2014user} have shown that compared to normal users, depressed users tend to post pictures with a relatively colder color. Therefore, we extracted three hue and saturation-related features to distinguish depressed users from normal users.

However, the warmth and coolness of a picture is a relative concept, and the human eye will give different conclusions when contrasting different colors. Lin et al. \cite{lin2014user} proposed a range definition of cold colors by analyzing the hue rings, which is used as our definition of the cold color range as $h_\mu \in(30, 110]$.

For the three color-related features, we compute them using values from the Hue, Saturation, Value (HSV) color model. Similarly to Red-Yellow-Blow (RGB) color space, HSV is a color space that represents the intuitive properties of colors, which is composed of hue, saturation, and lightness. Among the three attributes, ``hue'' refers to the category of colored light. Different wavelengths of light give different colors and hues. It is measured by the angle value, with a range of 0-360 degree. From red to counterclockwise, the red hue is defined as 0 degree, the green is 120 degree, and the blue is 240 degree.

Saturation indicates the degree of color close to the color of the spectrum, and usually takes a value of 0 to 1. The larger the value, the more saturated the color. After converting the RGB value of each pixel to the HSV color space, we calculate the dominant color ternary $\mu=(h_\mu, s_\mu, v_\mu)$. The algorithm for extracting the dominant color is given in Algorithm 1.
\begin{center}
    \begin{minipage}[htbp]{12cm}
    \begin{algorithm}[H]
        \caption{Dominant Color Extraction}
        \LinesNumbered
        \KwIn{$\tau$, All the pixels of a picture, represented in the HSV color space.}
        \KwOut{The dominant color pixel of the picture}

        \textbf{Initialize: } $threshold \gets 30$ \par
        \tcc{the striking pixel threshold}
        \textbf{Initialize: } Array $SP\_Arr$ \par
        \tcc{to storage the striking pixels}
        $\overline{\tau} \gets$ the average of $\tau $
        \par
        \For{every $pixel$ \textbf{in} $\tau$}{
            $h_\tau=pixel[0]$ \tcp*{$pixel=(h_\tau, s_\tau, v_\tau)$}
            $\overline{h_\mu}=\overline{\tau}[0]$ \tcp*{$\overline{\tau}=(\overline{h_\mu}, \overline{s_\mu}, \overline{v_\mu})$} \par
            \uIf {$\mid h_\tau - \overline{h_\mu} \mid > threshold $}{
                $SP\_Arr \gets SP\_Arr + pixel$
            }
        }
        $\mu \gets $ the average of $SP$
        \par
        \Return $\mu$
        \label{algo_color}
    \end{algorithm}
    \end{minipage}
\end{center}

The dominant color is the most attractive and the dominant color in a picture. Thus, we introduce the striking pixel (SP) to represent these colors. The SP plays an important role in the intuitive perception of the entire picture, usually measured by the absolute difference between a specific pixel and the average hue of the entire picture. The algorithm first inputs a picture with all the pixels represented by the HSV color space. It initializes a manually assigned threshold and an array $SP\_Arr$ to store striking pixels. Then, the algorithm calculates the average color $\overline{\Omega}=(\overline{h_\mu}, \overline{s_\mu}, \overline{v_\mu})$ of the picture and iterates through each pixel, comparing the absolute value of the difference between its hue value and $\overline{h_\mu}$. If the difference is greater than the threshold, the currently iterated pixel is defined as a SP. Finally, by calculating the average value of the SP array, the dominant color ternary $\mu$ is calculated. Several rounds of tests have been ran to choose the best value of the threshold (here set to 30).

We count the total number of posted pictures with $h_\mu \in (30, 110]$ and $s_\mu < 0.7$ as $C_{(\pi_{cold})}$. Then, $\gamma_{PCP}$ is calculated by:
\begin{equation}\label{gamma_pci}
    \gamma_{PCP} = \frac{1}{C_{(\pi)}} \times C_{(\pi_{cold})}, \gamma_{FPP} \in [0,1]
\end{equation}

\textbf{Standard deviation of hue} and \textbf{Standard deviation of saturation}. These two features are used to reflect the fluctuation of the user's picture color. The previous works used the mean values of hue and saturation as the picture features and achieved good results on Twitter \cite{shen2018cross,lin2014user} and Weibo \cite{shen2017depression,shen2018cross,lin2014user}. In our research, we found that the hue of depressed users' pictures is more concentrated in colder ranges and the saturation value is relatively low. On the contrary, the hue and saturation distribution of normal users is more dispersed and average. We take the hue value $h\mu$ and the saturation value $s_\mu$ to calculate their mean values $\overline{X}_{SDH}$ and $\overline{X}_{SDS}$ by:
\begin{equation}\label{x_sdh}
    \overline{X}_{SDH}=\frac{1}{C_{(\pi)}} \times \sum_{i=1}^{C_{(\pi)}}h_{\mu_i}
\end{equation}
\begin{equation}\label{x_sds}
    \overline{X}_{SDS}=\frac{1}{C_{(\pi)}} \times \sum_{i=1}^{C_{(\pi)}}s_{\mu_i}
\end{equation}

Then, $\gamma_{SDH}$ and $\gamma_{SDS}$ can be defined using the following equations:
\begin{equation}\label{gamma_sdh}
    \gamma_{SDH}=\sqrt{\frac{1}{C_{(\pi)}} \times \sum_{i=1}^{C_{(\pi)}}(h_{\mu_i}-\overline{X}_{SDH})^{2}}
\end{equation}
\begin{equation}\label{gamma_sds}
    \gamma_{SDS}=\sqrt{\frac{1}{C_{(\pi)}} \times \sum_{i=1}^{C_{(\pi)}}(s_{\mu_i}-\overline{X}_{SDH})^{2}}
\end{equation}

\subsection{Word Feature Extracting}
\subsubsection{Text sequence construction}
Algorithm 2 gives the approach to construct the user text sequence. Considering that the user nickname and profile can also reflect its current emotional state, they are also concatenated to the tweet text.

\begin{center}
    \begin{minipage}[htbp]{12cm}
    \centering
    \begin{algorithm}[H]
        \caption{User Long Text Sequence Construction}
        \LinesNumbered
        \KwIn{$T$, a collection of user nicknames, profiles, and all tweets' text.}
        \KwOut{The concatenated long text sequence $S_\xi$}

        \textbf{Initialize:} An empty string $S_\xi$ \par
        \textbf{Initialize:} The max length of text sequence $\Delta$ \par
        \For{$text$ in $T$}{
            \uIf{the length of $S_\xi > \Delta$}{
                \textbf{break}
            }
            \uIf{$text$ belongs to an original post}{
                $S_\xi \gets S_\xi + text$
            }
            \uElseIf{$text$ belongs to a repost}{
                \uIf{$text$ = ``Repost''}{
                    \tcc{ignored the default repost reason}
                    \textbf{continue}
                }
                \uElse{
                    $S_\xi \gets S_\xi + text$
                }
            }
            \uElse{
                \tcp{User nickname or profile}
                $S_\xi \gets S_\xi + text$
            }
        }
        \Return $S_\xi$
        \label{algo_seq}
    \end{algorithm}
    \end{minipage}
\end{center}

The algorithm firstly inputs all the text information (defined as $T$) of the user and constructs the concatenated user long text sequence $S_\xi$ by traversing $T$. After entering the loop, the algorithm first determines if the current length of the concatenated text sequence is greater than the maximum length $\Delta$; the algorithm ends if the condition is satisfied. Then, it concatenates the user's original tweet text in chronological order from the latest to the earliest. Moreover, when a user repost a tweet, Weibo will ask the user to fill in the reason for the retweet. In particular, if the user does not fill in the reason, the text ``\begin{CJK*}{UTF8}{gbsn}转发微博\end{CJK*}'' (``Repost'' in English) will be automatically added as default. This default repost reason is not retained in the text sequence $S_\xi$ since it does not express any opinions and feelings.

\subsubsection{Word Embedding}
To effectively vectorize the text sequence constructed above and apply this feature to the classification algorithm, the characteristics of this long text sequence are further discussed.

First, the sequence is strongly contextually linked. This link exists not only within a single tweet but also among the contexts of multiple tweets. For example, a user posts multiple tweets at different times about depression diagnoses, depression onset, medication treatment, and inner distress. The integration of these information points is usually the key to judge whether a user is depressed.

Secondly, considering that under real circumstances, not all the tweets would describe depression-related content even for true depression users. That is, capturing text semantics such as ``the user states that he has been diagnosed with depression'' and ``the user expresses a strong inclination of suicide'' is critical for detection depression using the long text sequence $S_\xi$.

Considering these aspects, several state-of-the-art word embedding algorithms are discussed here. Transformer \cite{vaswani2017attention} is a model that replaces the recurrent neural network with the attention mechanism. It calculates the weights of each unit in a long sequence to effectively capture the important semantic information. Moreover, BERT \cite{devlin2018bert} is a two-way encoder that is proposed recently. However, due to the limitations of the ``Position Embedding'' structure in BERT (including its derivative models ALBERT and ROBERTa), the maximum sequence length $Delta$ for single processing is restricted to 512 units. Furthermore, the existing truncation or batch processing of long text sequences used in Ref. \cite{lin2020sensemood} will significantly increase the time complexity of processing, which is considered as an unsuitable solution for performing a timely depression detection task. Therefore, the ideal word embedding model must have the ability to process long text efficiently and accurately.

A novel language model, namely XLNet \cite{yang2019xlnet}, is then proposed by Yang et al. Since it combines the features of language models such as auto-regression and auto-encoding, XLNet has resolved the problem that BERT ignores the relationship between the Masked locations and can process longer text sequences. In this paper, XLNet-Chinese-base\footnote{https://github.com/ymcui/Chinese-XLNet} is used as the upstream word embedding model, then a multitask-based DNN classifier, FusionNet, is implemented to handle the downstream tasks.

\subsection{Multitask Learning Classification Model: FusionNet}
\begin{figure*}[htbp]
    \centering
    \includegraphics[scale=0.32]{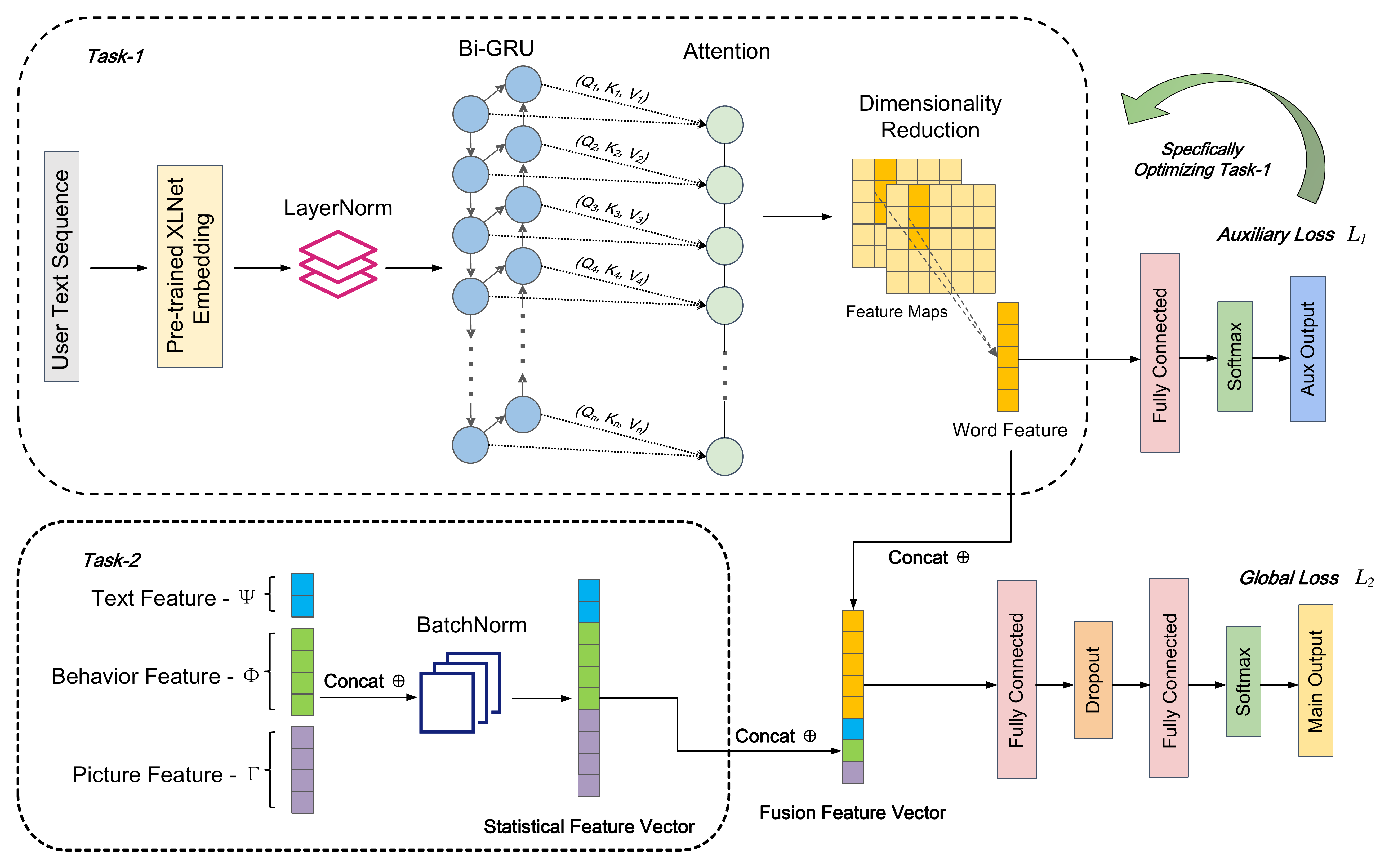}
    \caption{The Structure of our Proposed FusionNet (FN)}
    \label{fusionnet}
\end{figure*}
Multitask learning is an integrated learning strategy that synchronizes model training in a way that multiple tasks share collective network structures and weights. Based on multitask learning, we construct a DNN classifier with Bi-GRU with attention as its main structure. As shown in Fig. \ref{fusionnet}, the word vector classification task \emph{(Task-1)} obtained from the upstream embedding model XLNet and the manually extracted statistical feature classification task \emph{(Task-2)} are considered as two classification tasks for detecting depressed users. Loss functions $L_1$ and $L_2$ with different weights $\omega_1$ and $\omega_2$ are manually defined to simultaneously train and optimize the network.

Firstly, the user text sequence $S_\xi$ is embedded by XLNet, and the output of the last hidden layer is connected to the layer normalization (LN) \cite{ba2016layer} layer. Then, the LN layer is connected to the Bi-GRU layer with attention to capture the key information and reduce the dimensionality of the word vector.

For \emph{Task-1}, this one-dimensional word feature connects the Full Connected (FC) layer, the Dropout layer, and the Softmax layer to directly output classification results. We set a auxiliary loss function $L_1$ for network optimization in` \emph{Task-1}, to help accelerate its network convergence. For \emph{Task-2}, the word feature is concatenated to the manually extracted statistical feature input. The statistical feature groups $[\Psi,\Phi,\Gamma]$ are regularized by the Batch Normalization (BN) layer \cite{ioffe2015batch}.

Moreover, the fused feature vector accesses multiple FC layers with activation functions, activating the hidden layers' outputs to further improve the fitting capability of the network. Finally, the network is connected to the Softmax layer, and the classification result is given by the main output. The main loss function $L_2$ is used to optimize the whole FusionNet network.

We define the weight parameter set of \emph{Task-1} network as $\Theta_{aux}$, while the objective function is defined as $\hat{f}_1$. The global weight parameter set of the whole network is represented by $\Theta_{g}$, while the objective function is defined as $\hat{f}_2$, so that $\Theta_{aux} \in \Theta_{g}$. Adopting the multitask learning stratergy, the joint optimization function $J$ can be described as:
\begin{equation}
    \label{eq_loss_j1_j2}
    \left\{ \begin{array}{lcl}
            J_1 = \sum_{i = 1}^{C_{(U)}} L_1(y_i, {\hat{f}}_1({S_\xi}_i), \Theta_{aux}), \\
            J_2=\sum_{i = 1}^{C_{(U)}} L_2(y_i, {\hat{f}_2}{(\Psi_i, \Phi_i, \Gamma_i), \Theta_g}).
            \end{array} \right.
\end{equation}

\begin{equation}
    \label{eq_joint_loss}
    J_{(\Theta_{aux}, \Theta_g)} = \omega_1 \times J_1+\omega_2 \times J_2
\end{equation}

In equation (\ref{eq_loss_j1_j2}), $y_i$ represents the true label (normal or depressed) of a specific user sample $i$. The $\hat{f}_1$ and the $\hat{f}_2$ both output the predicted label of user sample $i$. In equation (\ref{eq_joint_loss}), $\omega_1$ and $\omega_2$ are the manually assigned weights of loss function $L_1$ and $L_2$.

Manually extracted features and several uncertain parameters will be evaluated in the following section.

\section{Experimental Evaluation and Analysis}
\subsection{Experiment Setup}
\subsubsection{Data cleaning}
Since the original user data obtained by the crawler has irrelevant information, to minimize the experimental bias and improve the efficiency of model training, we have removed all the non-text contents in the tweets.

\subsubsection{Dataset slicing}
In this part, WU3D is divided into four subsets: $D_1$, $D_2$, $D_3$, and $D_4$. All of the subsets are sampled using a fixed random seed without a crossover.

Among them, $D_1$ is used for DNN model training and the 10-fold cross-validation of TML classifiers. Furthermore, $D_2$ is used as a fixed dataset for validation in each round of neural network training. Finally, we test the models' performance on $D_3$ and give the evaluation metrics. As a supplementary dataset, $D_4$ contains only 325 depressed users and 12,245 normal users, which will be only used in the last experiment of unbalanced training samples. Statistics of the sliced datasets are given in Table \ref{data_part}:

\begin{table}[htbp]
    \centering
    \caption{Dataset Slicing Statistics}
    \begin{tabular}{ccccccc}
        \hline
        \multirow{2}{*}{\textbf{Dataset}} & \multicolumn{3}{c}{\textbf{Depressed}} & \multicolumn{3}{c}{\textbf{Normal}} \\ \cline{2-7}
             & \textbf{User} & Tweet  & Picture & \textbf{User} & Tweet   & Picture \\ \hline
        WU3D & 10325         & 408797 & 160481  & 22245         & 1783113 & 1087556  \\
        $D_1$   & 8000          & 319115 & 125096  & 8,000          & 630064  & 355353  \\
        $D_2$   & 1000          & 37315  & 14991   & 1000          & 79417   & 46060   \\
        $D_3$   & 1000          & 38941  & 15211   & 1000          & 80066   & 45066   \\
        $D_4$   & 325           & 13426  & 5183    & 12245          & 993566  & 641077  \\ \hline
    \end{tabular}
    \label{data_part}
\end{table}

\subsubsection{Evaluation metrics}
The experimental metrics used in this section mainly includes supervised machine learning metrics. \textbf{True Positive (TP), True Negative (TN), False Positive (FP), False Negative (FN)} are commonly used to describe the number of classes predicted by models in classification tasks. Among them, TP represents the number of depressed users correctly predicted, TN represents the number of normal users correctly predicted, FP represents the number of depressed users incorrectly predicted, and FN represents the number of normal users incorrectly predicted. Based on the above four metrics, we can further define the advanced metrics by:
\begin{equation}\label{accuracy}
  Accuracy=\frac{\mid TP+TN \mid}{\mid TP+TN+FP+FN \mid}
\end{equation}
\begin{equation}\label{precision}
  Precision=\frac{\mid TP \mid}{\mid TP+FP \mid}
\end{equation}
\begin{equation}\label{recall}
  Recall=\frac{\mid TP+TN \mid}{\mid TP+FN \mid}
\end{equation}
\begin{equation}\label{f1}
  F_1-Score=\frac{2 \times Precision \times Recall}{Precision+Recall}
\end{equation}

Moreover, Receiver Operating Characteristic (ROC) curve is a curve with False Positive Rate (FPR) as the horizontal axis and True Positive Rate (TPR) as the vertical axis, which can be used to visually reflect the classifier performance. Furthermore, in the experiment of the statistical features, we introduce the Man-Whitney U test and the cumulative distribution function (CDF) curve as the evaluation metrics.

\subsubsection{Training settings}
\textbf{(i) Hardware and software environments.} We complete all the experiments in this section on a workstation with Intel Xeon Silver 4212 CPU, NVIDIA RTX TITAN GPU with 24GB GRAM, and 32GB RAM. The programming-related settings used in the experiments are Python v3.7.5, Anaconda v4.8.3, TensorFlow v2.1.0, and Scikit-learn v0.23.1.

\textbf{(ii) Baseline statistical feature classifiers.} For the statistical feature classification task, we select several popular TML model from existing studies to demonstrate the effectiveness of the ten concluded features.
\begin{itemize}
    \item \textbf{LR:} Logistic Regression is a commonly used linear model \cite{cohan2018smhd} and has good classification performance.
    \item \textbf{NB:} A Naive Bayesian classifier is a simple probabilistic classifier using Bayes' theorem as a basis. Its implementation is relatively simple and is used more often in related works \cite{wang2013depression,deshpande2017depression,al2019depression,shen2017depression,lin2014user,balani2015detecting}.
    \item \textbf{SVM:} Support Vector Machine classifiers apply the principle of structural risk minimization to the field of classification and are the most used classifiers in previous studies \cite{de2013predicting,deshpande2017depression,stankevich2018feature,al2019depression,mustafa2020multiclass,orabi2018deep,sadeque2018measuring,lin2014user,cheng2017assessing}. In our work, we discussed different kernel modes including the linear kernel, polynomial kernel and radial basis kernel of the SVM, respectively.
    \item \textbf{RF:} Random Forest is an algorithm that integrates multiple trees through ensemble learning, which is also used widely in related works \cite{stankevich2018feature,mustafa2020multiclass,lin2014user,balani2015detecting}. The basic unit of RF is the decision tree.
    \item \textbf{AB:} Adaptive Boosting is an ensemble learning algorithm that combines multiple simple classifiers \cite{cohan2018smhd}.
    \item \textbf{GBDT:} Gradient Boosting Decesion Tree is a classification model that uses an integrated additive model to continuously reduce the training residuals. GBDT is one of the algorithms with an excellent generalization ability in TML, however, to the best of our knowledge, there is no existing work using GBDT as a classification model.
    \item \textbf{BP:} The Back Propagation (BP) network is extracted from the main output part of our proposed FusionNet with the same parameter settings. To be specific, BP is composed of ``FC+Dropout+FC+Softmax''.
\end{itemize}

\textbf{(iii) Baseline word vector classification networks.} For word vector classifiers, we use several popular neural network structures as the main structures, appended by the FC layers and the Softmax layer to output the classification label.
\begin{itemize}
    \item \textbf{CNN-1D:} One-dimensional convolutional neural networks are more widely used in natural language processing, and have achieved good performance in the task of depression detection \cite{mustafa2020multiclass,orabi2018deep,trotzek2018utilizing,lin2014user,cohan2018smhd}.
    \item \textbf{Bi-LSTM:} The bi-directional LSTM network splices two-way LSTMs together, which are more capable of handling time series data \cite{orabi2018deep,sadeque2018measuring,lin2020sensemood}.
    \item \textbf{Bi-GRU:} Similarly, the bi-directional GRU splices the two-directional GRU network together and is similar to Bi-LSTM in its ability to handle time series data \cite{sadeque2018measuring,gui2019cooperative}.
    \item \textbf{TCN:} The temporal convolutional network is a new algorithm for processing time series that reduces the serial processing complexity of RNNs \cite{pandey2019tcnn}.
    \item \textbf{Attention:} The attention mechanism is proposed by Vaswani et al. \cite{vaswani2017attention}, which can quickly filter out high-value information from large amounts of information. Attention is popular in many fields such as machine translation and speech recognition.
    \item \textbf{Bi-GRU with attention:} It is extracted from our proposed FusionNet with the same parameter settings.
    \item \textbf{Bi-GRU with GAP:} Global Average Pooling (GAP) \cite{lin2013network} is used to replace the attention layer to reduce the dimensionality of the output of Bi-GRU, so as to compare the performance differences of these two similar structures.
\end{itemize}

For the baseline statistical feature classifiers and word vector classification networks, we have run a series of pre-experiments on each classifier and selected the structures and parameters with the best classification performance. Each classifier will be represented by its main structure's symbols (e.g., Bi-GRU-based classifiers are referred to as Bi-GRU for short). API default parameters are selected for both TML and DL classifiers that are not specifically described here. Moreover, separate experiments for the neural network structures of BP and Bi-GRU with attention are set to further demonstrate the superiority of FusionNet, which uses multitask learning to merge the two DNN structures.

The loss functions and callback settings for neural network training are given in Table \ref{tab_neural_classifier_training}:

\begin{table}[htbp]
    \centering
    \caption{Neural Network Training Setup}
    \begin{tabular}{cc}
        \hline
        \multicolumn{1}{c}{\textbf{Item}} & \multicolumn{1}{c}{\textbf{Setup}}                    \\ \hline
        Batch size                        & 32                                                    \\
        Epoch                             & 80                                                   \\
        Early Stopping                    & monitor='val\_acc', patience=10                       \\
        Check Point                       & monitor='val\_acc', mode='max'                        \\
        FN Loss Function $L_1$, $L_2$     & Categorical Crossentropy                              \\
        FN Optimizer (for $L_1$)          & NAdam (Init\_lr=3e-4)                                 \\
        FN Optimizer (for $L_2$)          & NAdam (Init\_lr=1e-3)                                 \\
        FN ${[}\omega_1, \omega_2{]}$                   & ${[}0.2, 1.0{]}$                                        \\ \hline
    \end{tabular}
    \label{tab_neural_classifier_training}
\end{table}

\subsection{Statistical Feature Analysis}
\subsubsection{Mann-Whitney U test}
In this part, we perform a non-parametric Mann-Whitney U test on each manually extracted feature. The result is shown in Table \ref{mwu_test}. Since the statistical variables Mann-Whitney U-value and Wilcoxon W-value can be transformed into each other, only the U-values are reserved in the table.
\begin{table}[htbp]
    \centering
    \caption{Mann-Whitney U Test Results}
    \begin{tabular}{lccc}
        \hline
        \textbf{Symbol} & \textbf{Mann-Whitney U} & \textbf{Significance} & \textbf{Decision} \\ \hline
        $\psi_{NP}$              & 118,195,683.0             & $p < 0.001$     & Reject $H_0$         \\
        $\psi_{FDW}$              & 111,603,040.5             & $p < 0.001$     & Reject $H_0$         \\ \hline
        $\phi_{POP}$              & 71,425,809.5              & $p < 0.001$     & Reject $H_0$         \\
        $\phi_{PLNP}$              & 87,347,143.5              & $p < 0.001$     & Reject $H_0$         \\
        $\phi_{PF}$              & 19,999,341.5              & $p < 0.001$     & Reject $H_0$         \\
        $\phi_{SDPT}$              & 85,482,386.5              & $p < 0.001$     & Reject $H_0$         \\ \hline
        $\phi_{FPP}$              & 47,524,863.5              & $p < 0.001$     & Reject $H_0$         \\
        $\phi_{PCP}$              & 58,070,802.0              & $p < 0.001$     & Reject $H_0$         \\
        $\phi_{SDH}$              & 52,041,246.5              & $p < 0.001$     & Reject $H_0$         \\
        $\phi_{SDS}$              & 54,667,566.5              & $p < 0.001$     & Reject $H_0$         \\ \hline
    \end{tabular}
    \label{mwu_test}
\end{table}

In the experiment, the default null hypothesis $H_0$ is set to ``the distribution of this feature is the same on normal users and depressed ones''. At the 95\% confidence interval, $p<0.05$ will reject the null hypothesis, which is to admit the significant difference between normal and depressed users. The result shows that the $p$-value of each feature is less than 0.001 in the significance of Mann Whitney's bilateral test. Thus, it is concluded that all the features pass the test and have significant differences in the distribution of two types of users.

\subsubsection{Cumulative distribution analysis}
\begin{figure*}[htbp]   
    \centering
    \subfigure[$\psi_{NP}$]{
        \includegraphics[scale=0.20]{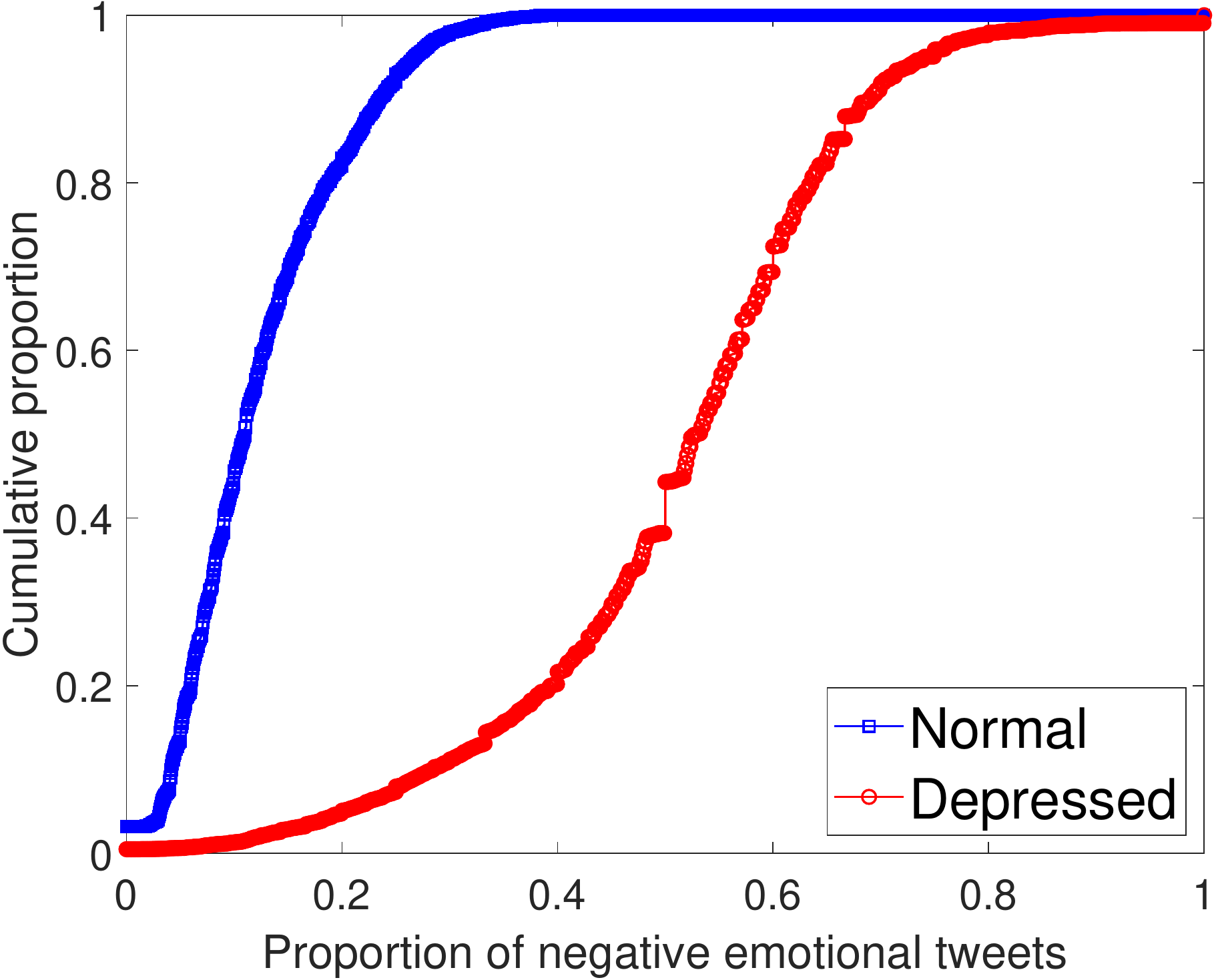}
    }
    \subfigure[$\psi_{FDW}$]{
        \includegraphics[scale=0.20]{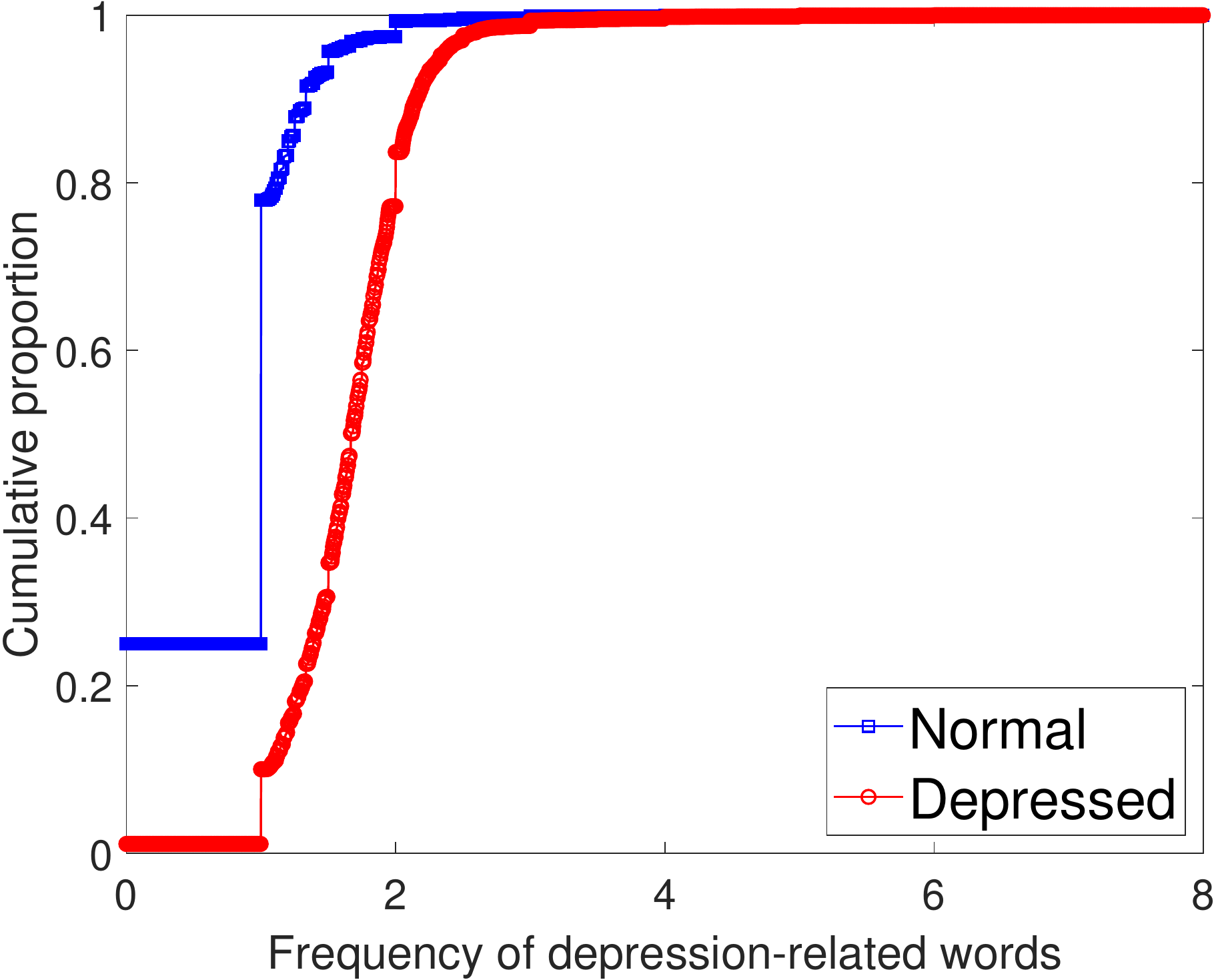}
    }
    \subfigure[$\phi_{POP}$]{
        \includegraphics[scale=0.20]{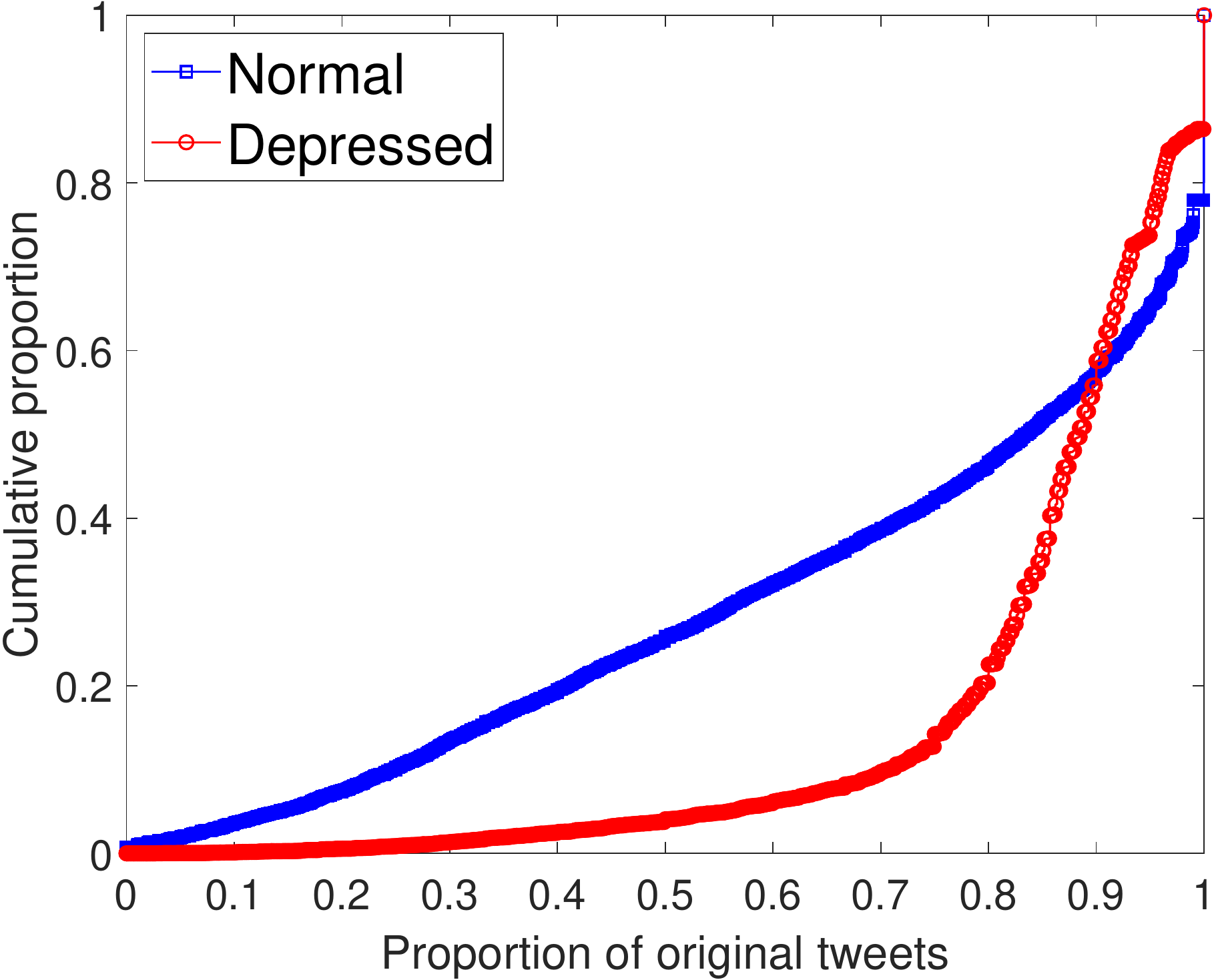}
    }   \par
    \subfigure[$\phi_{PLNP}$]{
        \includegraphics[scale=0.20]{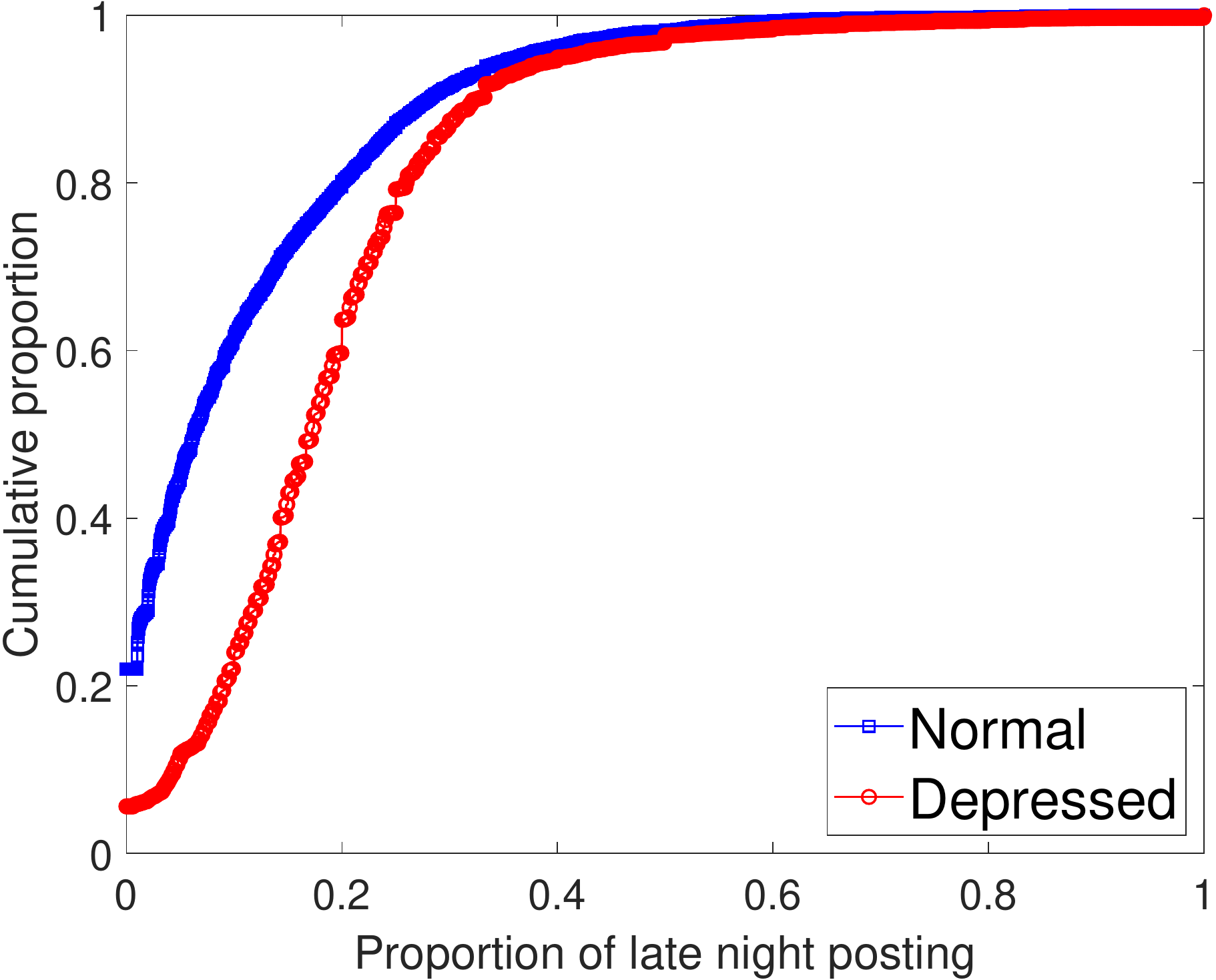}
    }
    \subfigure[$\phi_{PF}$]{
        \includegraphics[scale=0.20]{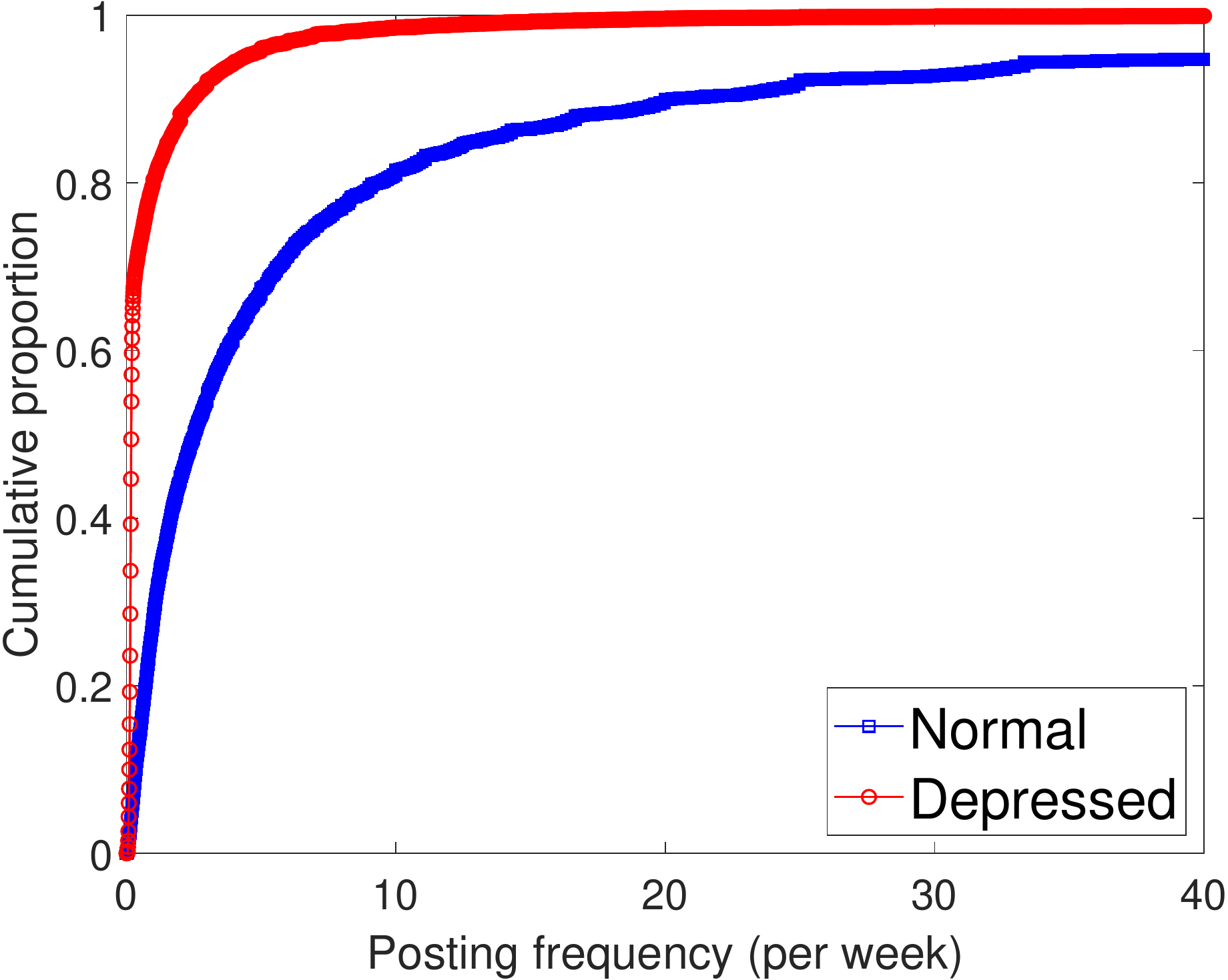}
    }
    \subfigure[$\phi_{SDPT}$]{
        \includegraphics[scale=0.20]{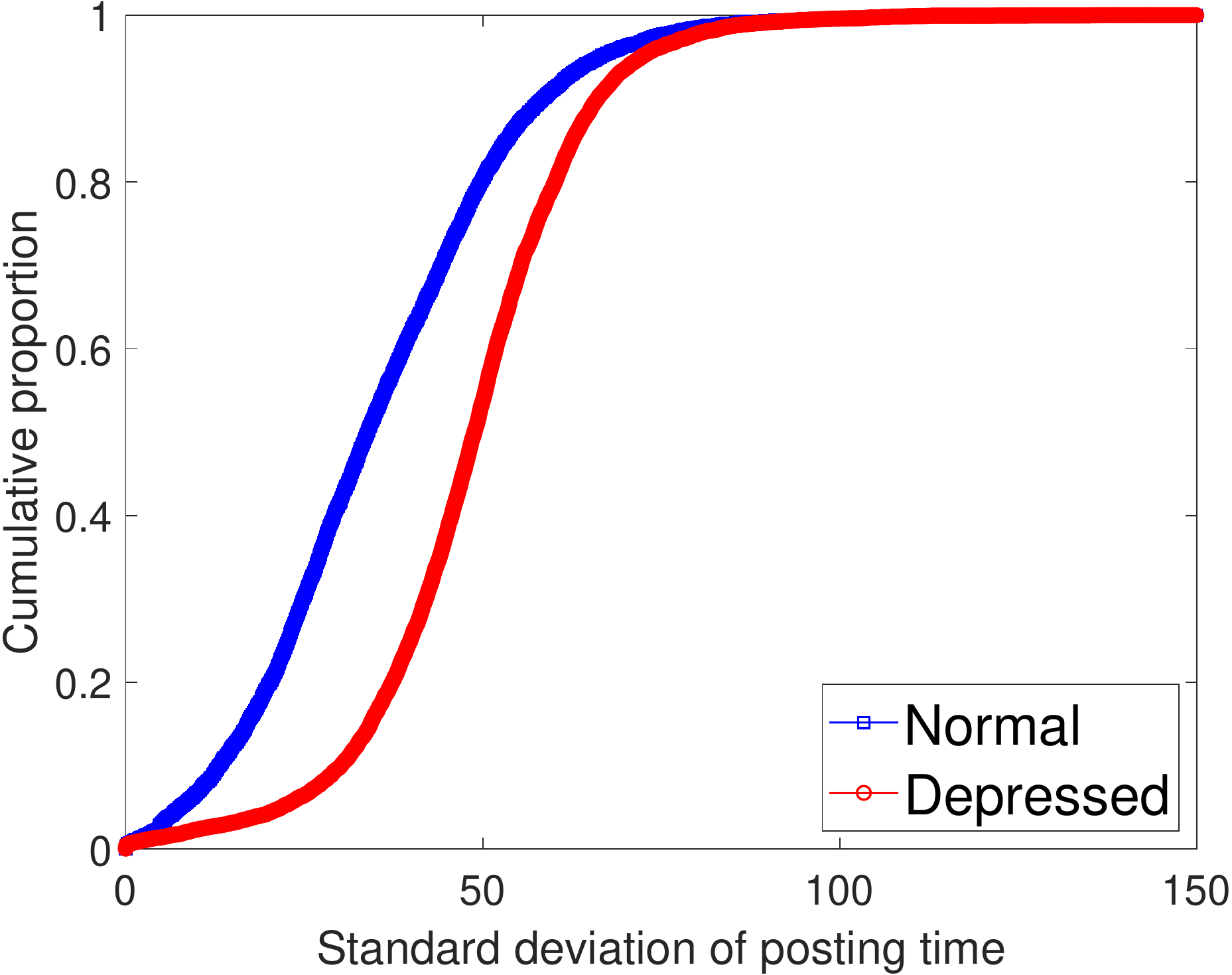}
    }   \par
    \subfigure[$\gamma_{FPP}$]{
        \includegraphics[scale=0.20]{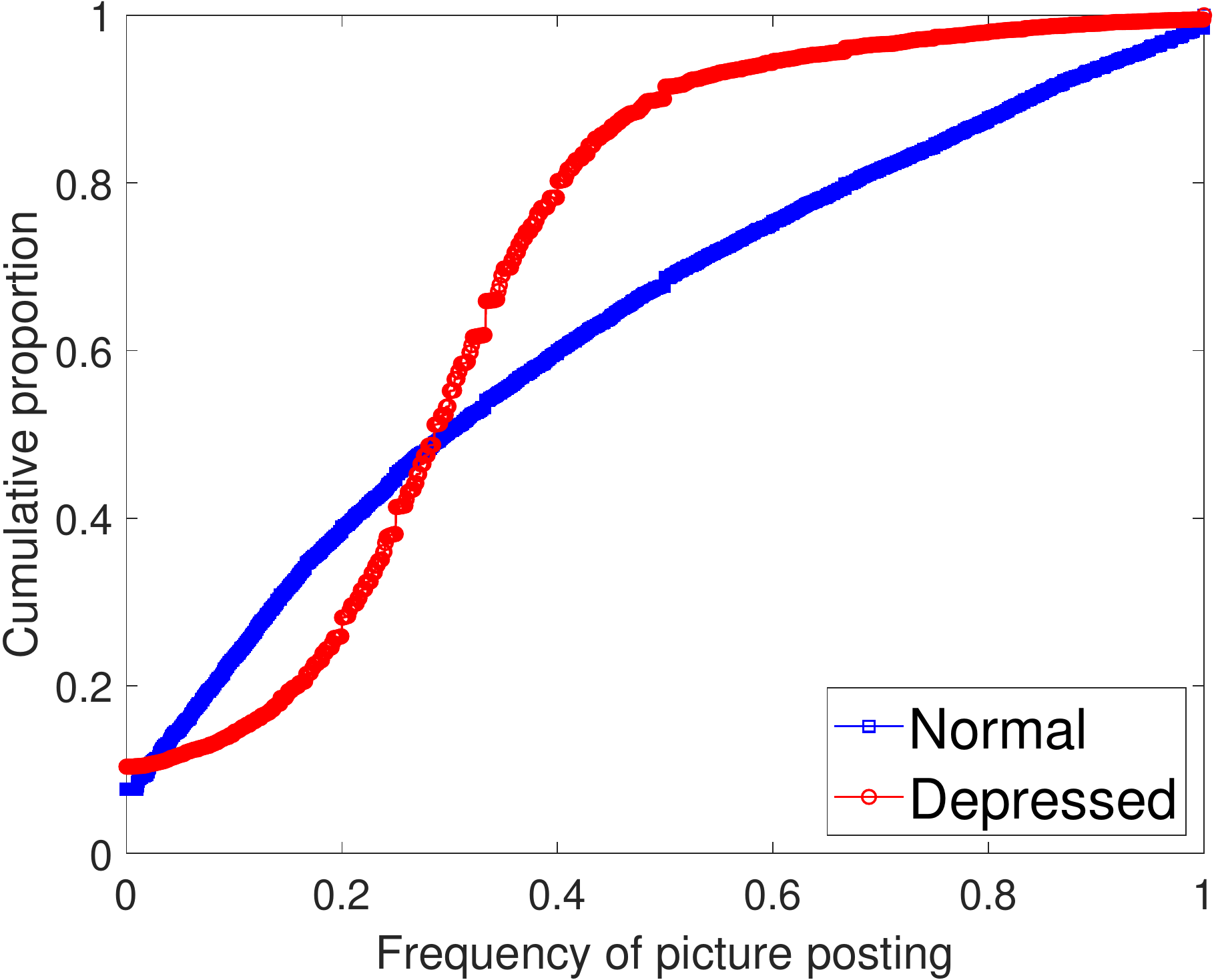}
    }
    \subfigure[$\gamma_{PCP}$]{
        \includegraphics[scale=0.20]{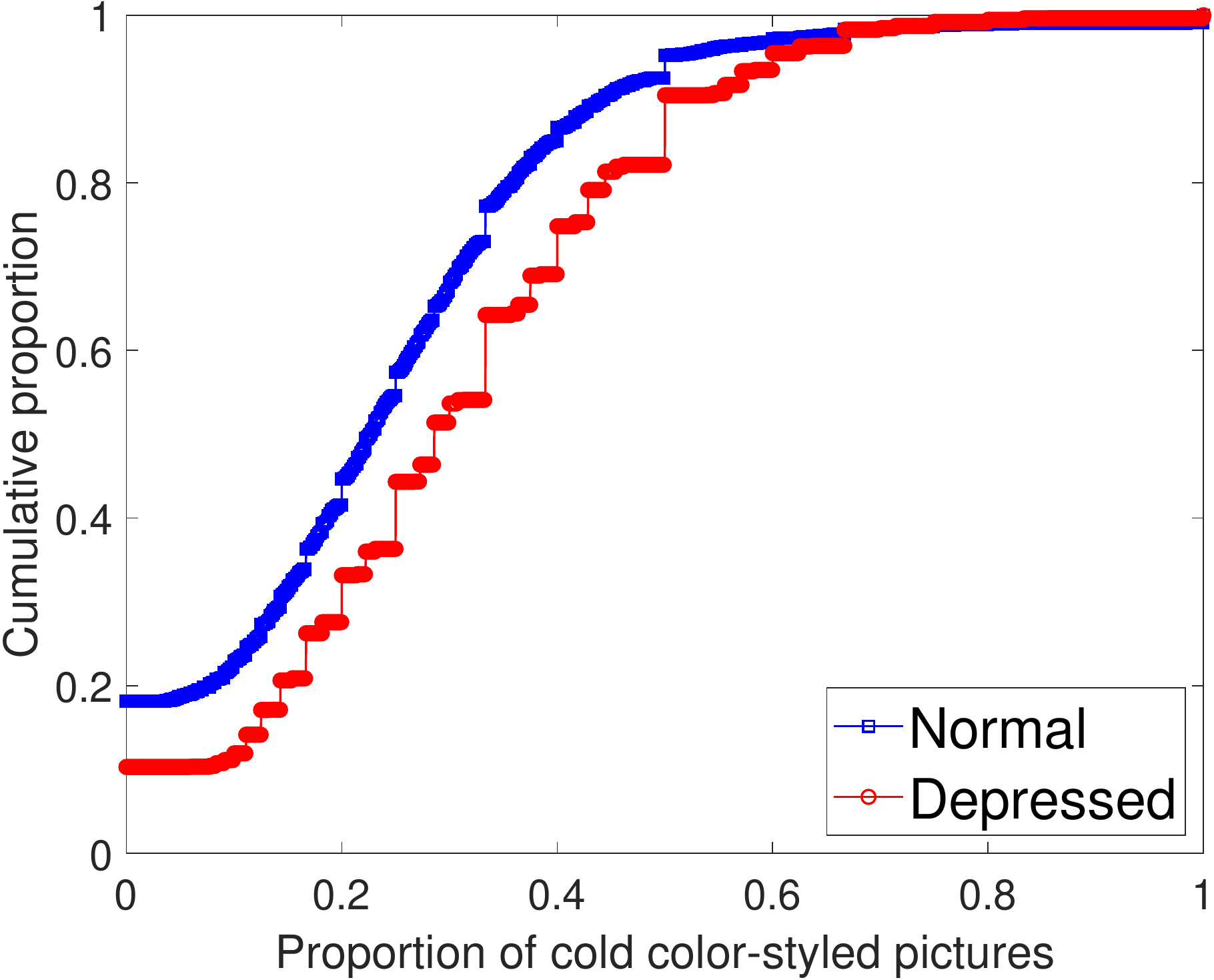}
    }
    \subfigure[$\gamma_{SDH}$]{
        \includegraphics[scale=0.20]{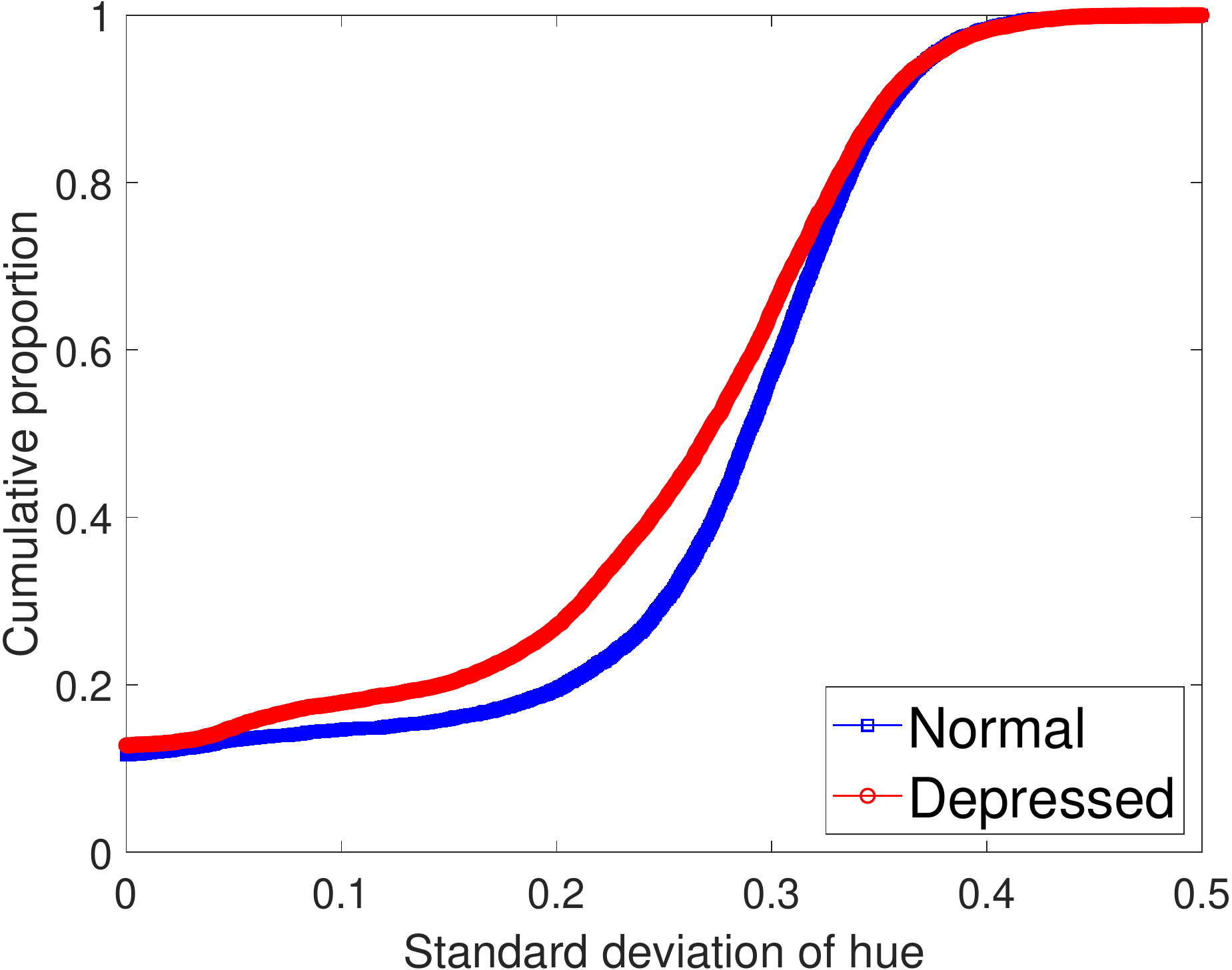}
    }
    \subfigure[$\gamma_{SDS}$]{
        \includegraphics[scale=0.20]{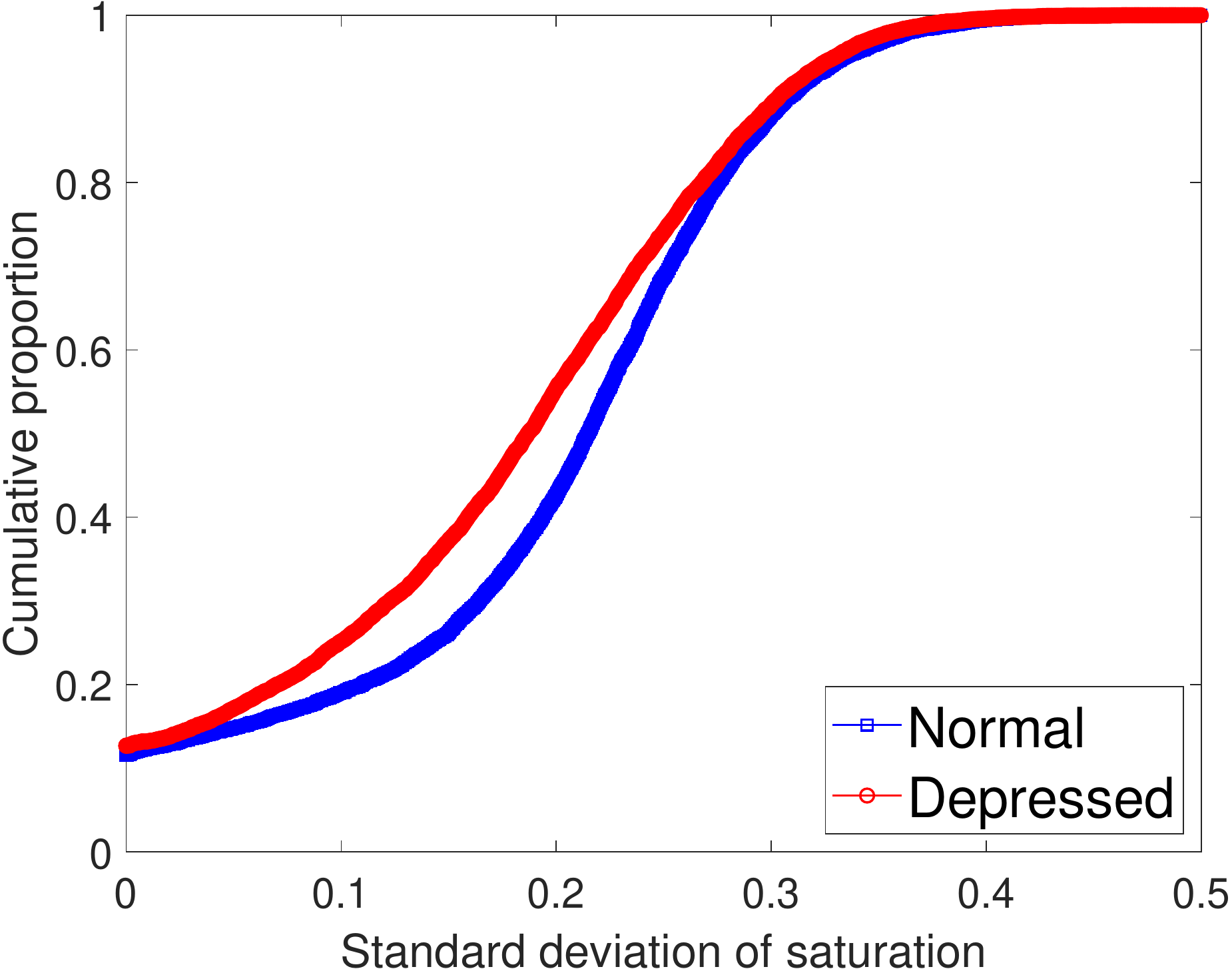}
    }
    \caption{The CDF Curves of Ten Statistical Features}
    \label{cdf-curve}
\end{figure*}

We also evaluate the significance of the features by comparing the feature distribution curve of normal and depressed users. The CDF curve for each feature are plotted in Fig. \ref{cdf-curve}. Since there are no quantitative parameters for the CDF curve to describe the results, we only evaluate the degree of coincidence between the two types of user curves.

Among the features, $\psi_{NP}$ has the highest distinction between the two types of users and the most significant difference in distribution. This phenomenon further demonstrates that text-based features play an important role in identifying depressed users on social networks. For our proposed features shown in Fig. \ref{cdf-curve}(a), (e), (f), and (g), the curve of two types of users shows obvious separation and different trends, indicating there are significant differences in these features between the two types of users.

\subsubsection{Classification Contribution of Different Feature Groups}
\begin{figure*}[htbp]
    \centering

    \subfigure[LR]{
        \includegraphics[scale=0.28]{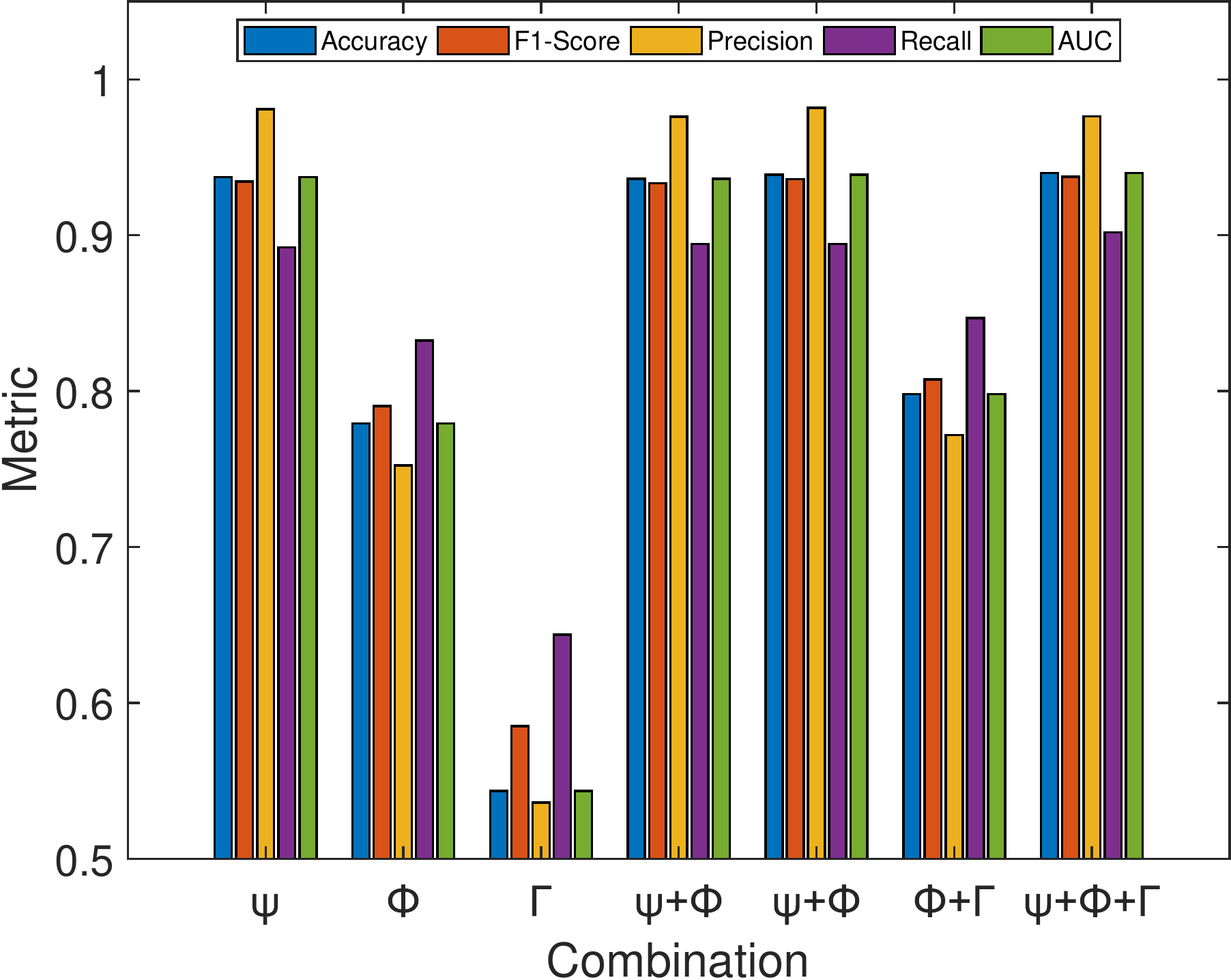}
    }
    \subfigure[NB]{
        \includegraphics[scale=0.28]{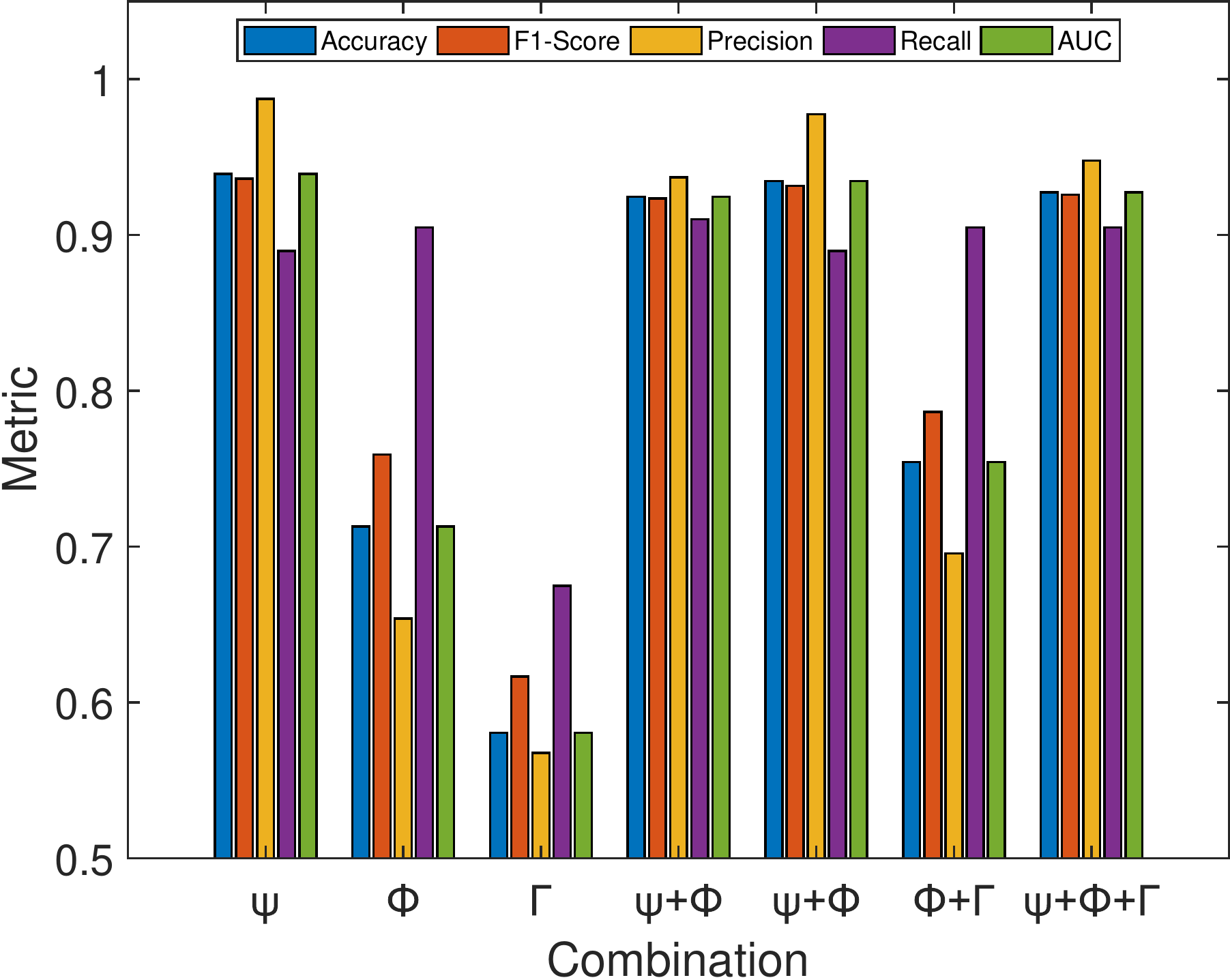}
    }
    \subfigure[SVM-linear]{
        \includegraphics[scale=0.28]{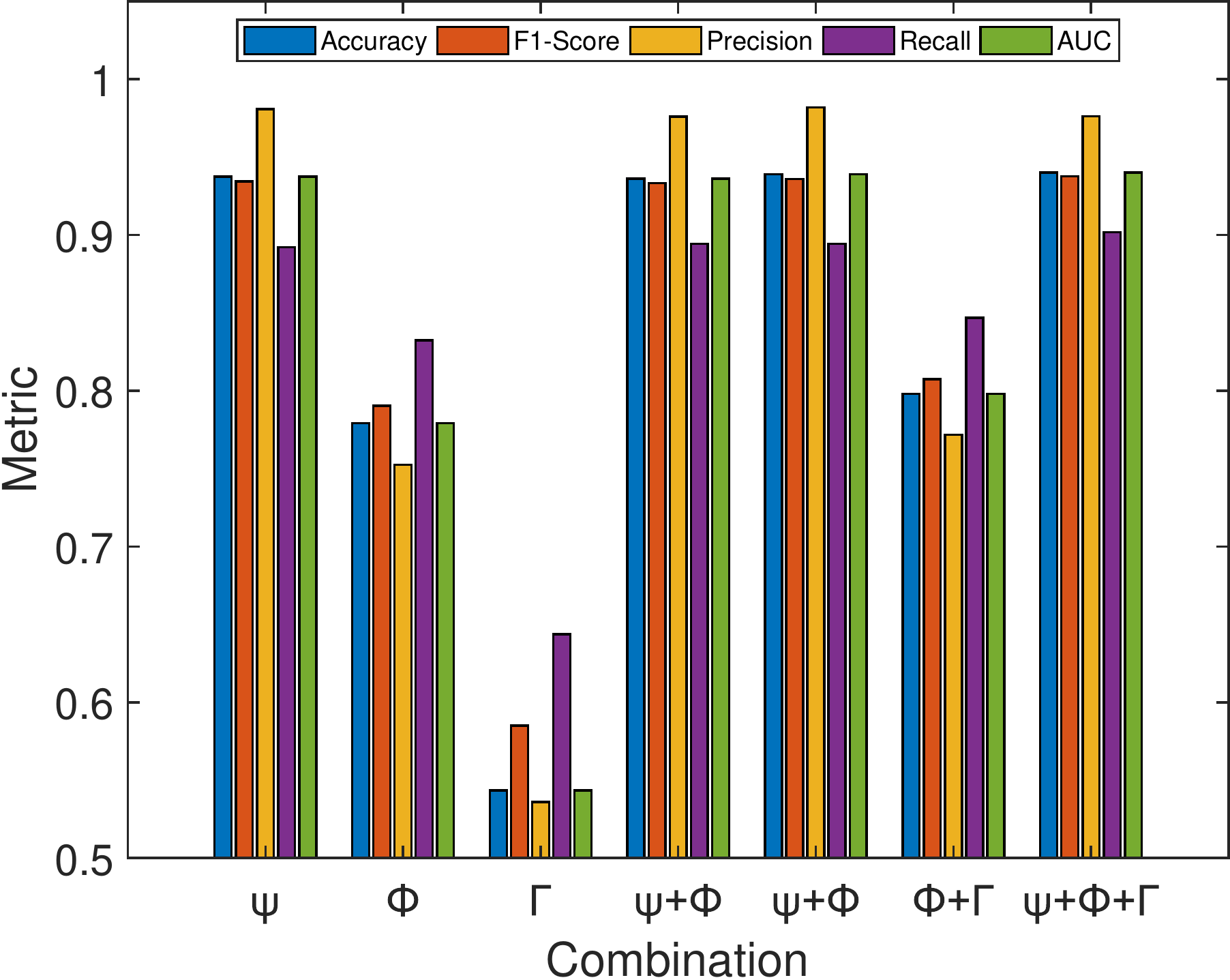}
    } \par
    \subfigure[SVM-poly]{
        \includegraphics[scale=0.28]{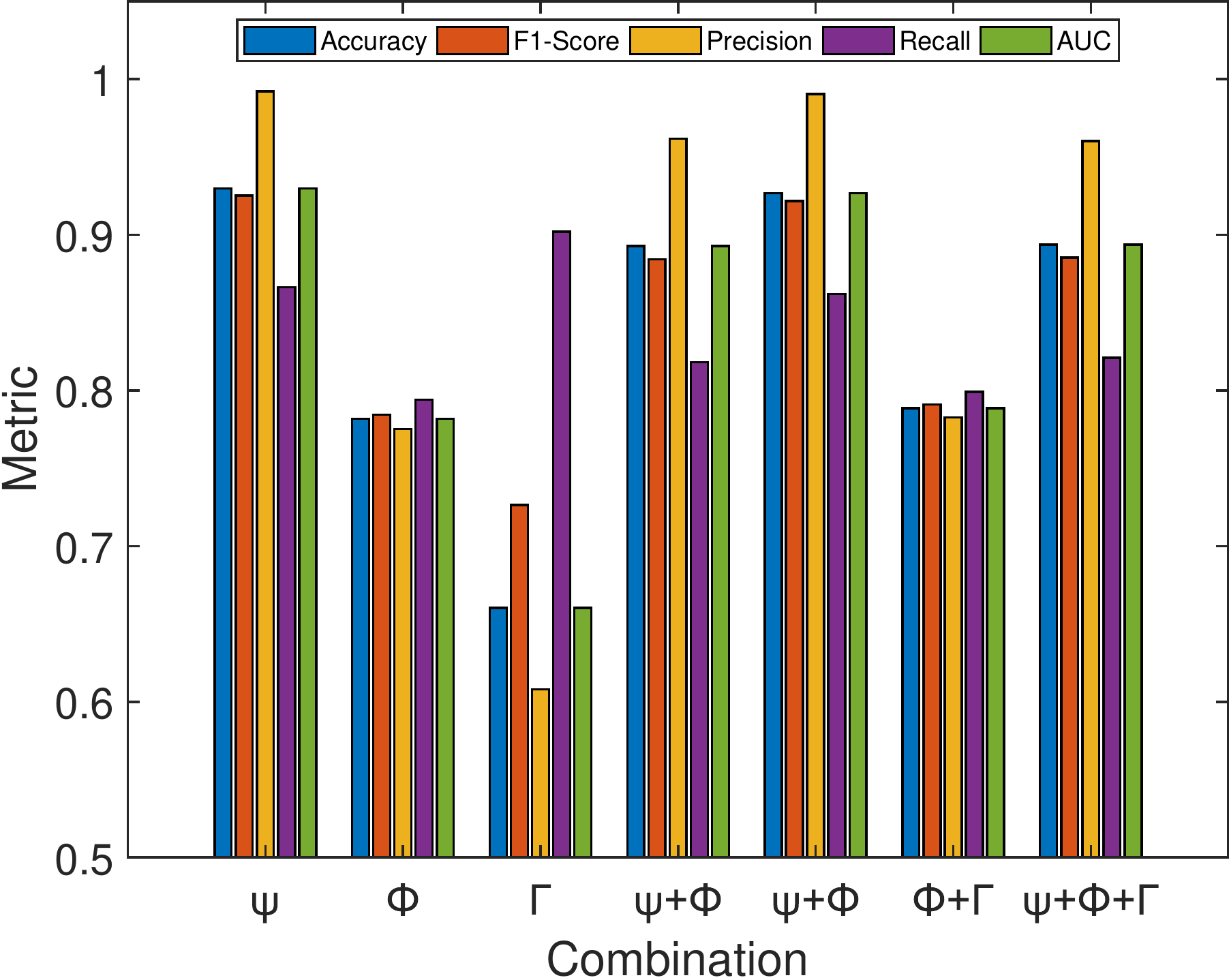}
    }
    \subfigure[SVM-rbf]{
        \includegraphics[scale=0.28]{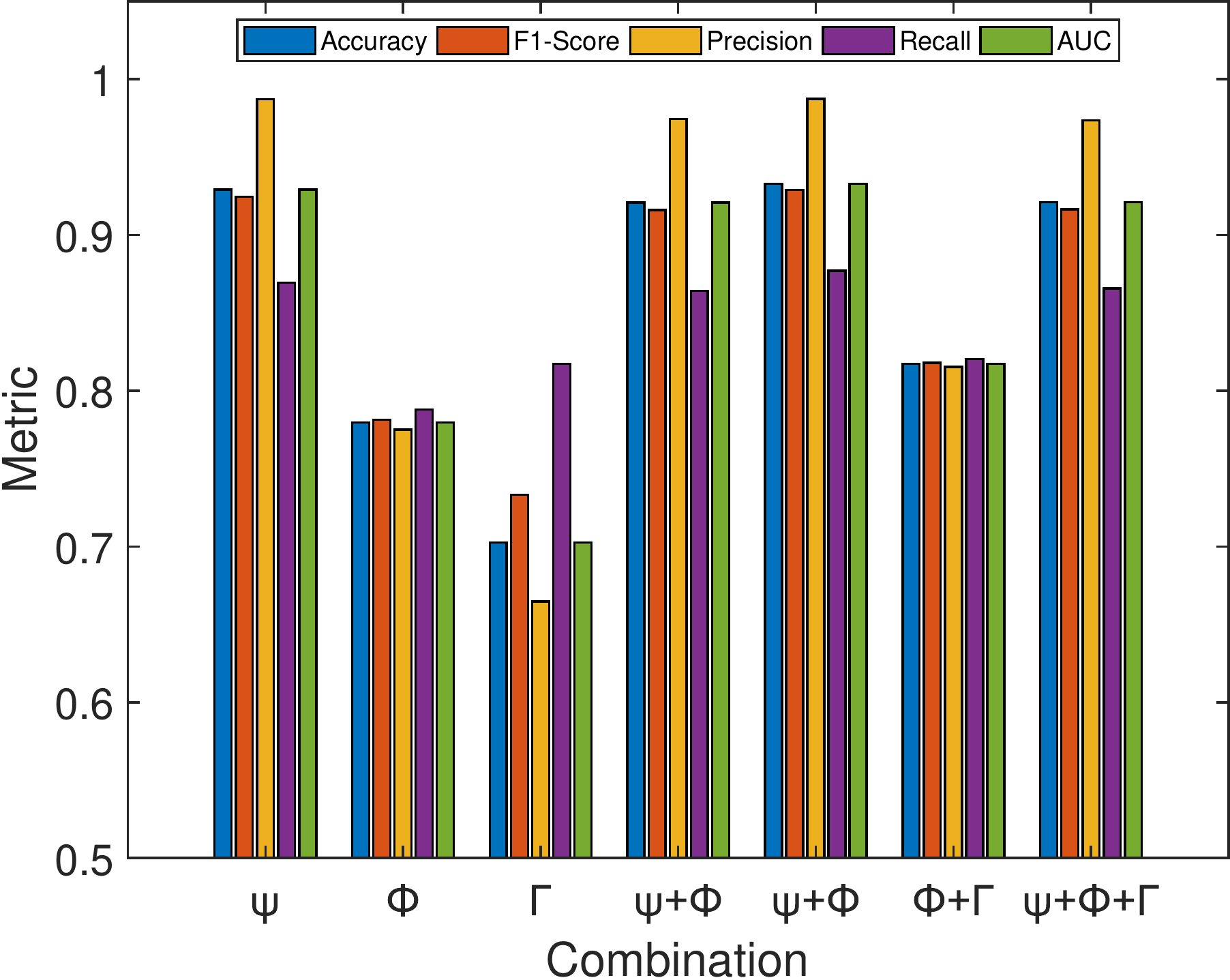}
    }
    \subfigure[RF]{
        \includegraphics[scale=0.28]{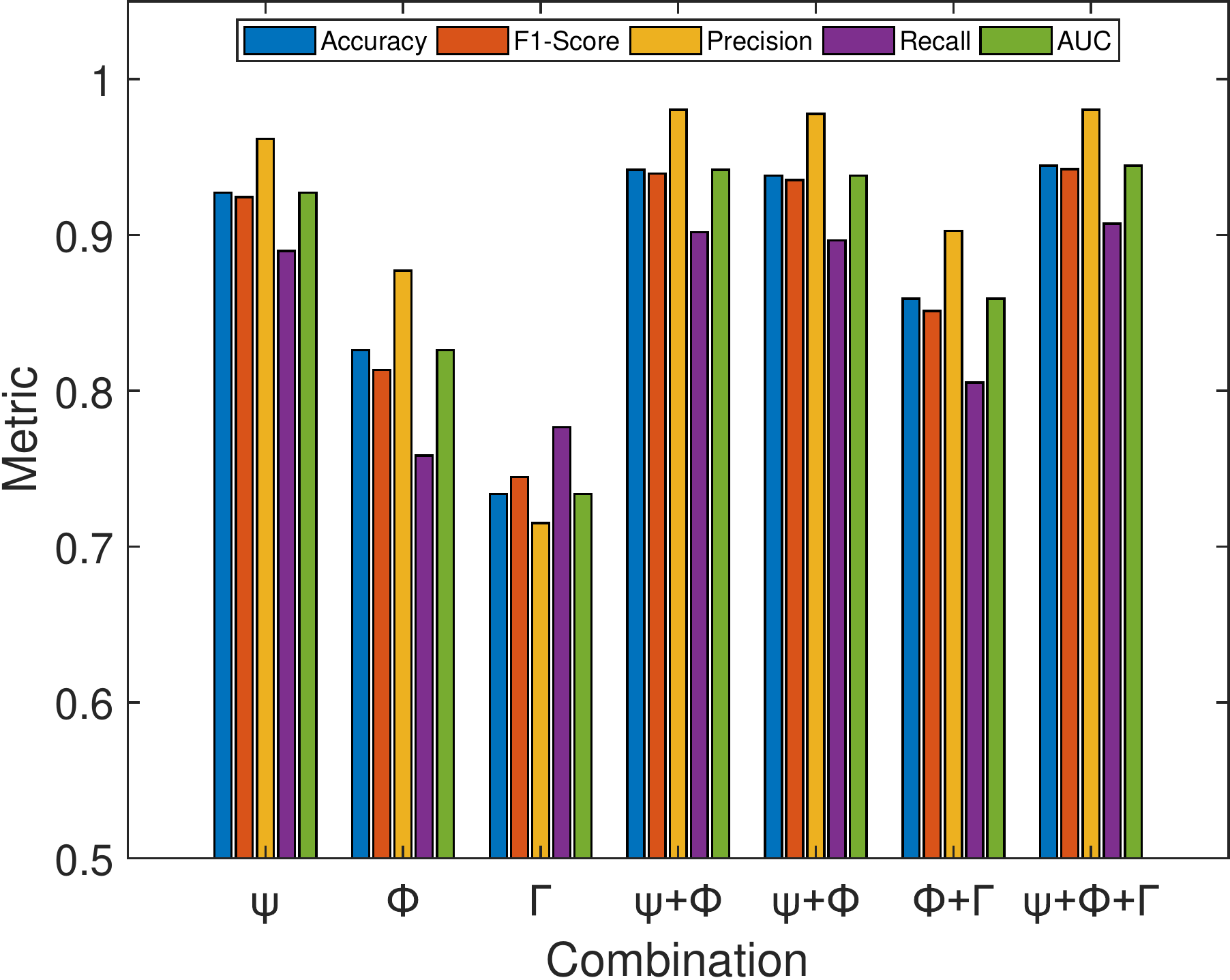}
    } \par
    \subfigure[AB]{
        \includegraphics[scale=0.28]{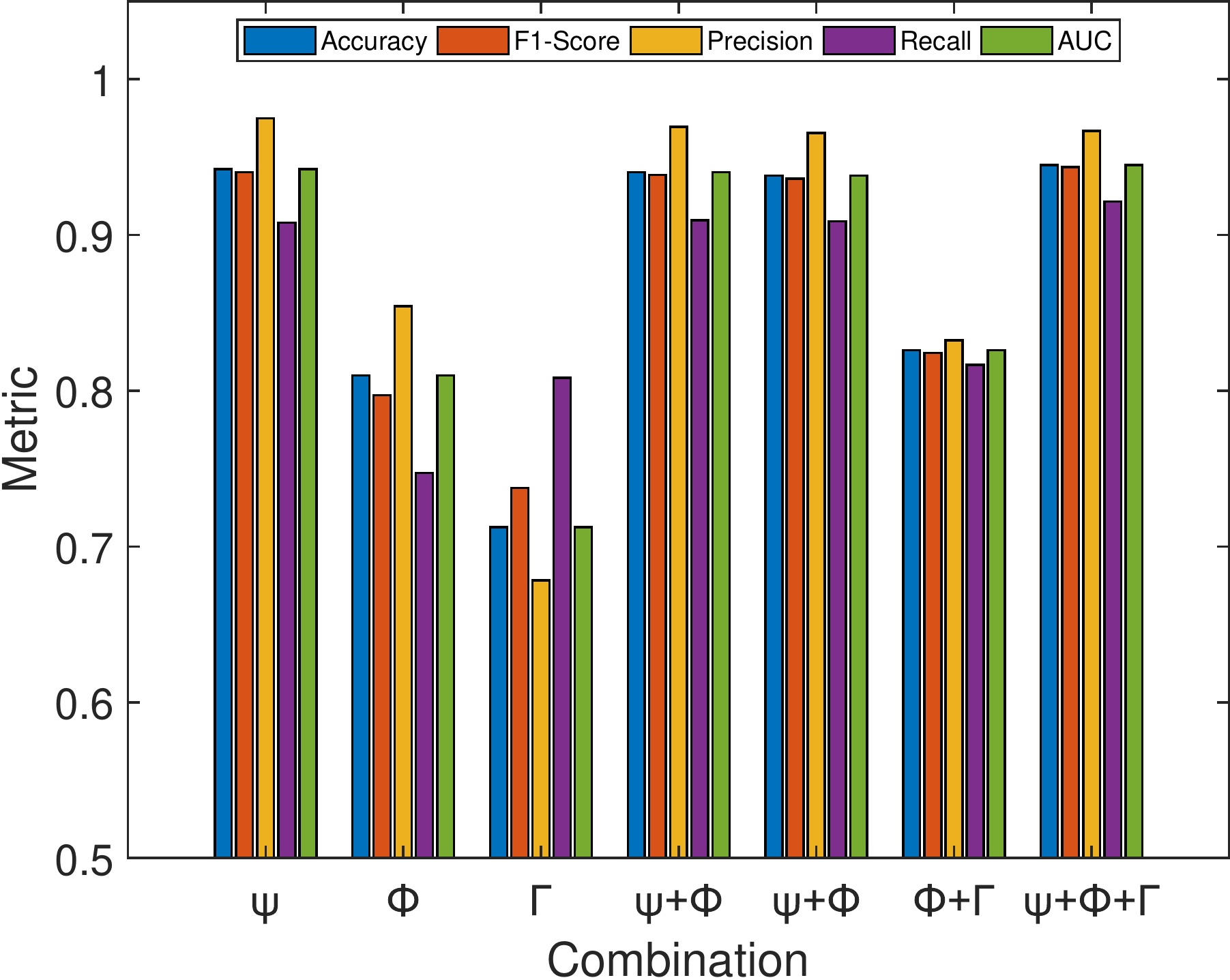}
    }
    \subfigure[GBDT]{
        \includegraphics[scale=0.28]{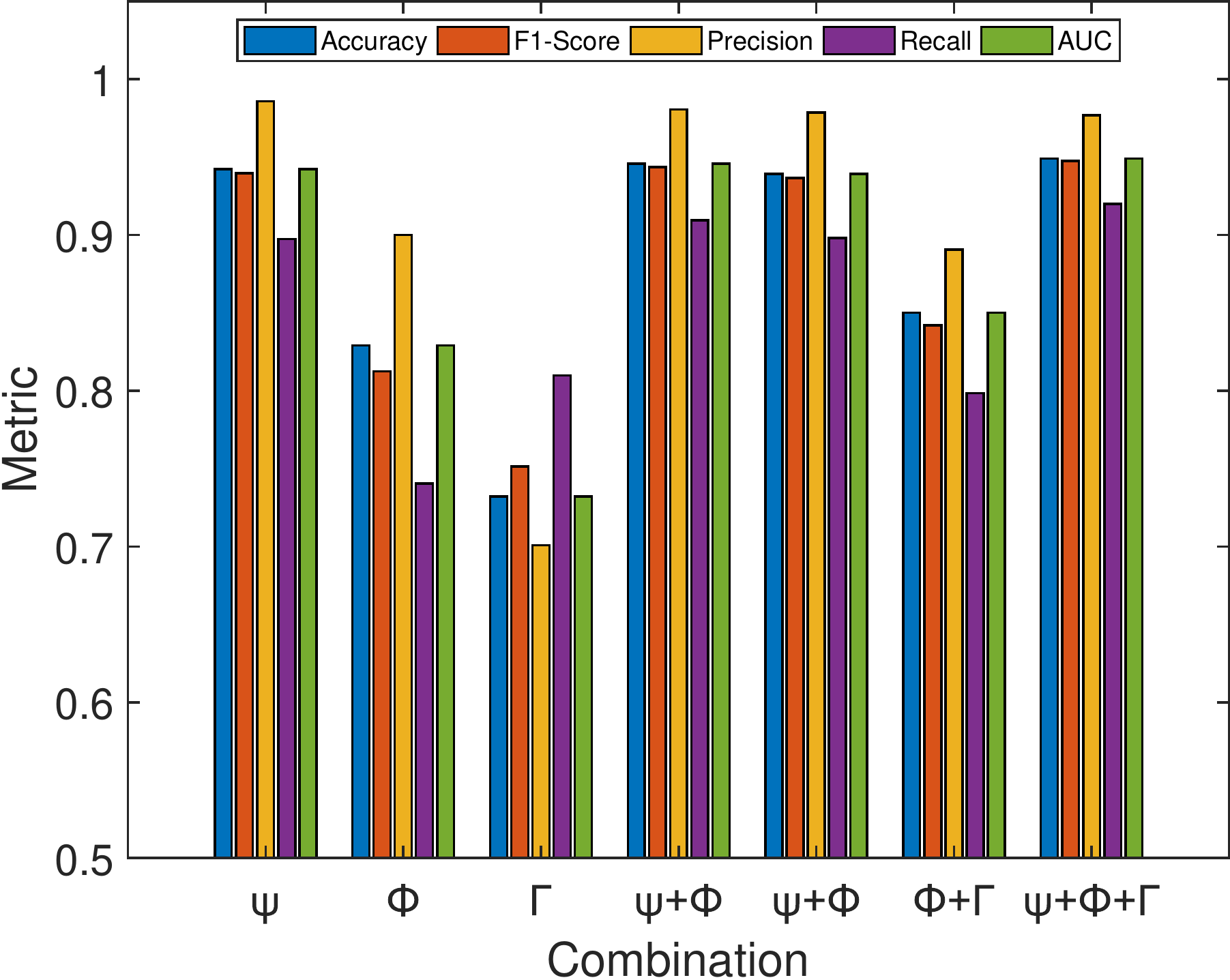}
    }
    \subfigure[BP]{
        \includegraphics[scale=0.28]{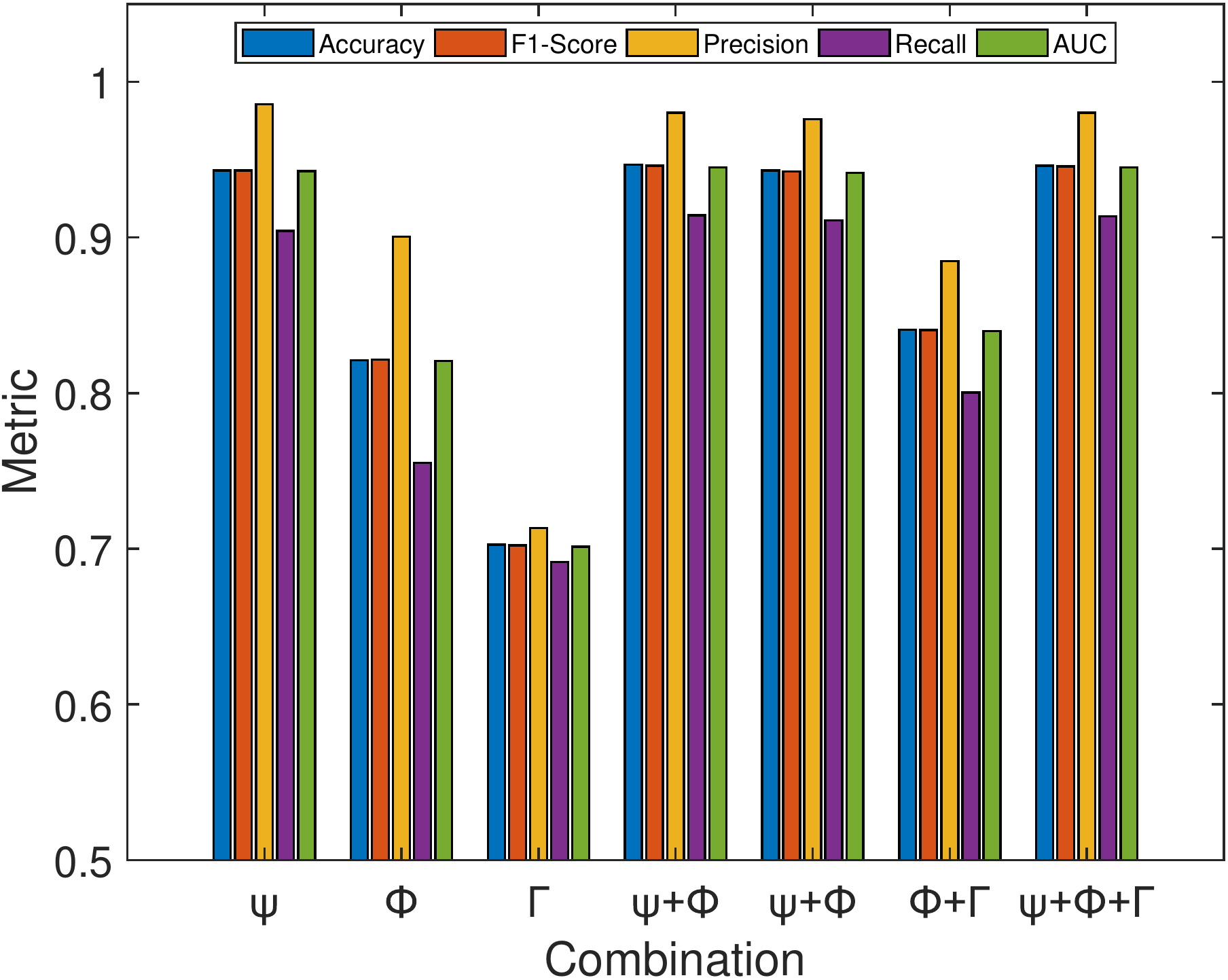}
    }

    \caption{The Contribution of Different Feature Group Combinations}
    \label{feature-contribution}
\end{figure*}

In this part, baseline statistical feature classifiers are used to evaluate the contribution of different feature groups. We perform experiments on different combinations of features to determine the contribution to the classification task. The result is shown in Fig. \ref{feature-contribution}.

The experimental result demonstrates that the classification performance of feature groups in each classifier is consistent with the results of our statistical tests, in which the text-based feature group $\Psi$ contributed the most. The BP classifier has already achieved a high F1-Score of 0.9431 when only text-based features are used. The contribution of the picture-based feature group $\Gamma$ is relatively poor, only with the highest F1-Score of 0.7514 using the GBDT classifier. However, under the different combinations of features, the performance is improved at different levels for each group and for each classifier. Especially, with the combination of all the feature groups ($\Psi+\Phi+\Gamma$), the GBDT classifier reports the highest F1-Score of 0.9465. GBDT and BP both achieved the highest performance metric for several rounds benefiting from gradient descent related optimization method.

Therefore, it is concluded that all three types of feature groups can positively improve the performance of classification tasks, with the text-based features contributing the most.

\subsection{Text Sequence Embedding Length Selection}
The user text information consists of the user nickname, profile, and tweet text, which are concatenated to a long text sequence using Algorithm 2. However, due to the uncertainty of the total number of tweets and the total number of words in a tweet, it is necessary to explore the most effective embedding length of the text sequence. In the meantime, since the pretrained Chinese XLNet has 12 layers of network structure, 768 hidden layers, and a total of 117M parameters, it will also take considerable time to access the network and to extract word vectors. Therefore, in this section, we run multiple experiments by setting several different values of the text sequence length. We record the time consumption of extracting word vectors using XLNet and the F1-Score of each classifier to explore the appropriate text sequence length. It should have reasonable embedding time consumption and relatively high F1-Score.

According to our statistics, a user's tweet text may generally be longer than 32 Chinese characters. A short text length may result in premature truncation of the tweet text. Thus, our experiment starts with sequence length $\Delta = 64$ and increases its value gradually. Finally, we select six groups of text sequence lengths for experiments, with $\Delta \in [64, 128, 256, 512, 1024, 2048]$. Then, we record the time consumption of embedding word vectors using XLNet and the classification F1-Score under different text sequence lengths.

\begin{figure}[htbp]
    \centering
    \subfigure[$\Delta$ - Time (Relative)]{
        \includegraphics[scale=0.4]{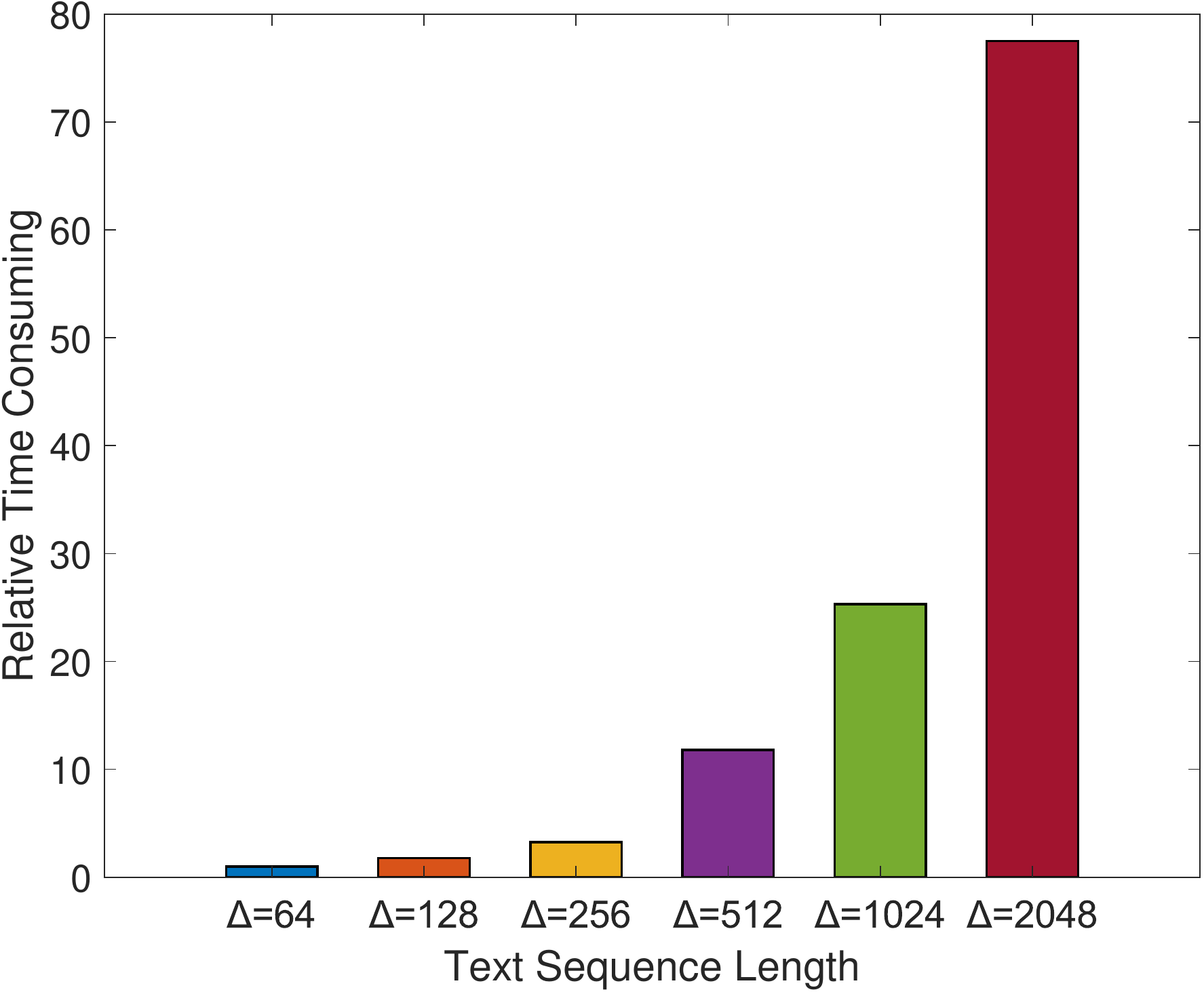}
    }
    \subfigure[$\Delta$ - F1-Score]{
        \includegraphics[scale=0.4]{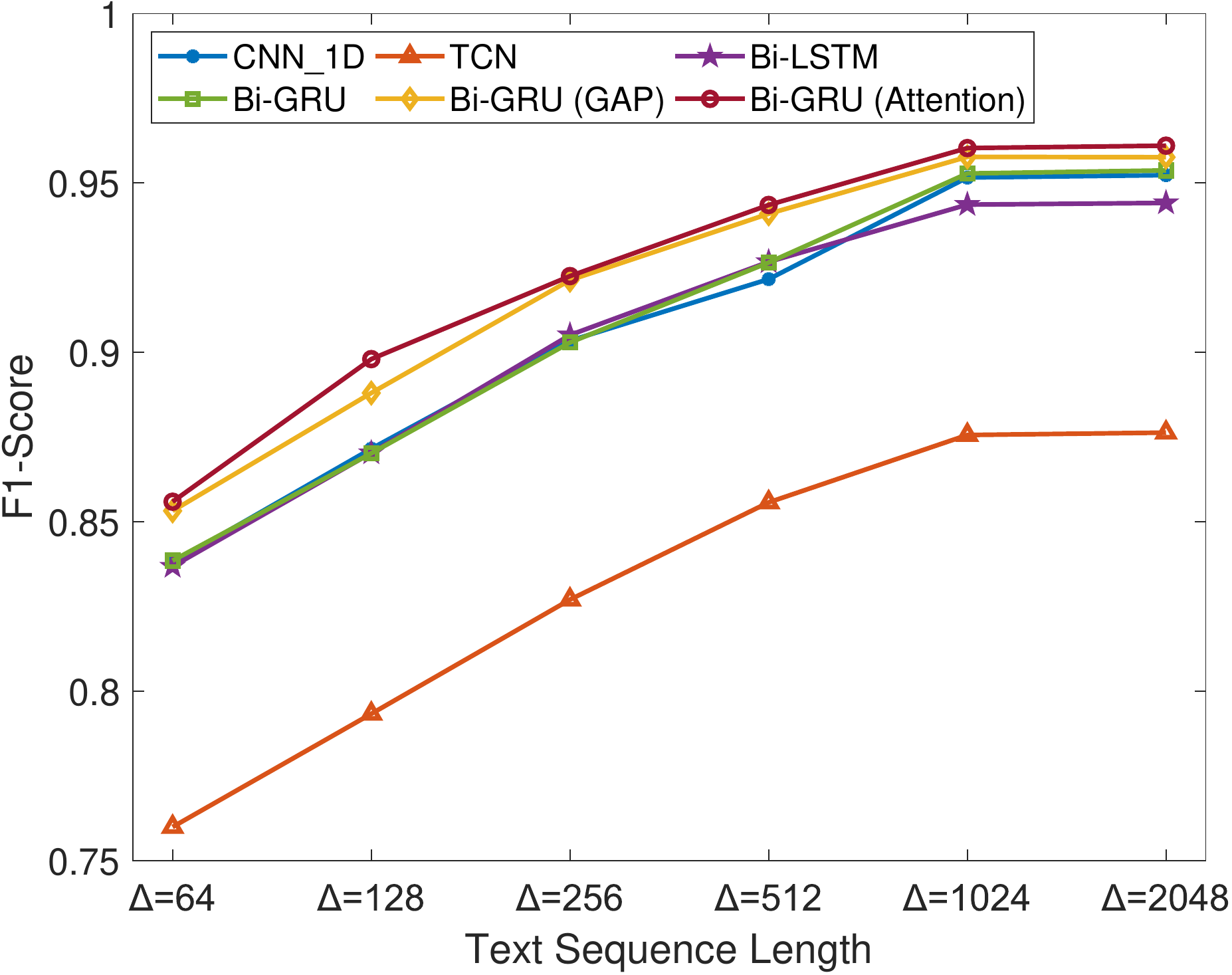}
    }
    \caption{The Selection of Text Sequence Length}
    \label{delta-time-f}
\end{figure}

We select the embedding time consumption of $\Delta=64$ as a base, normalizing the remaining groups to get the scaled values. Figure \ref{delta-time-f}(a) shows the relative time consumption of XLNet at each value of the text sequence length $\Delta$.

The result demonstrates that, as the text sequence length increases, the curve shows a nonlinear trend and the growth rate of the slope continues to increase. Specifically, the relative time of $\Delta=512$ is 11.81, while the corresponding time of $\Delta=1024$ is 25.31, and the corresponding time of $\Delta=2048$ is 77.52. The time consumption increases by 114.31\% and 206.28\%, respectively. Therefore, when $\Delta$ increases by a binary exponential power, the time of embedding will increase at a significantly faster rate.

Moreover, the baseline word vector classification networks are used to reflect the classification performance under different values of $\Delta$. Figure \ref{delta_time_f}(b) shows the F1-Score of each classifier under different $\Delta$ values.

When $\Delta$ increases from 64 to 1024, the F1-Score of each classifier increases significantly. When $\Delta$ increases from 1024 to 2048, the F1-Score almost no longer increases. Although the performance of a single attention structure is relatively poor, by adding it to other structures, the classification performance can be further improved. Thus, Bi-GRU with attention can better capture key information in long text sequences among the tested word vector classifiers. It achieves the highest classification performance and thus is used in our proposed FusionNet.

Combined with the previous experiment of $\Delta$ and the word vector extraction time consumption, it is concluded that the length of text sequence at $\Delta=1024$ has reached a performance bottleneck in this classification task, while the time computation of word embedding is relatively acceptable. Therefore, in subsequent experiments involving word vectors, we implement $\Delta=1024$ as the default text sequence embedding length.

\subsection{Performance of Target Classifiers}
To compare other classifiers with FusionNet, in this section, the output of Bi-GRU with attention is extracted as the word feature, which is concatenated to the statistical feature. Then, this integrated feature vector is input into the baseline statistical feature classifiers to accomplish the classification task of depressed and normal users.

\begin{figure*}[htbp]
    \centering
    \subfigure[Accuracy]{
        \includegraphics[scale=0.4]{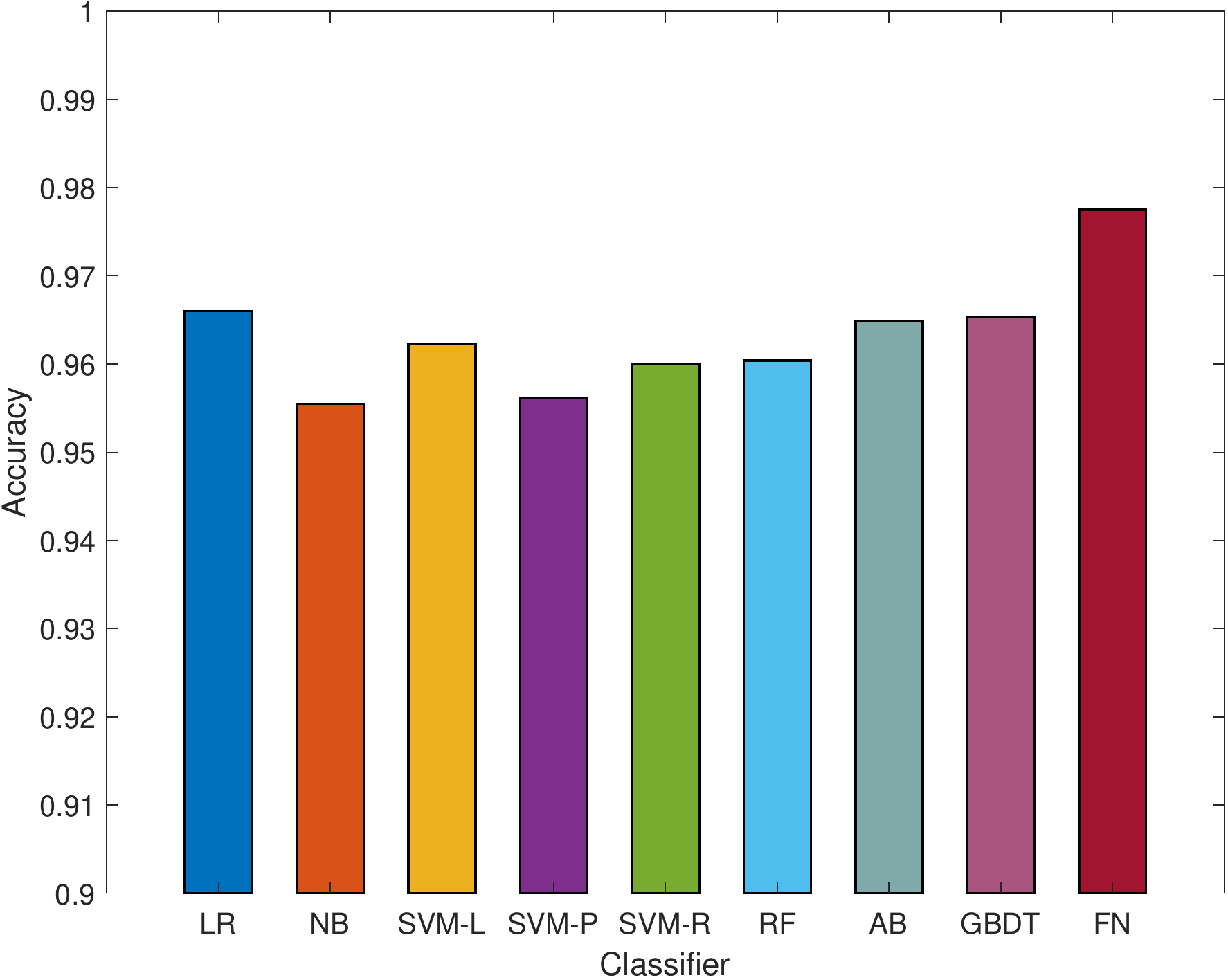}
    }
    \subfigure[F1-Score]{
        \includegraphics[scale=0.4]{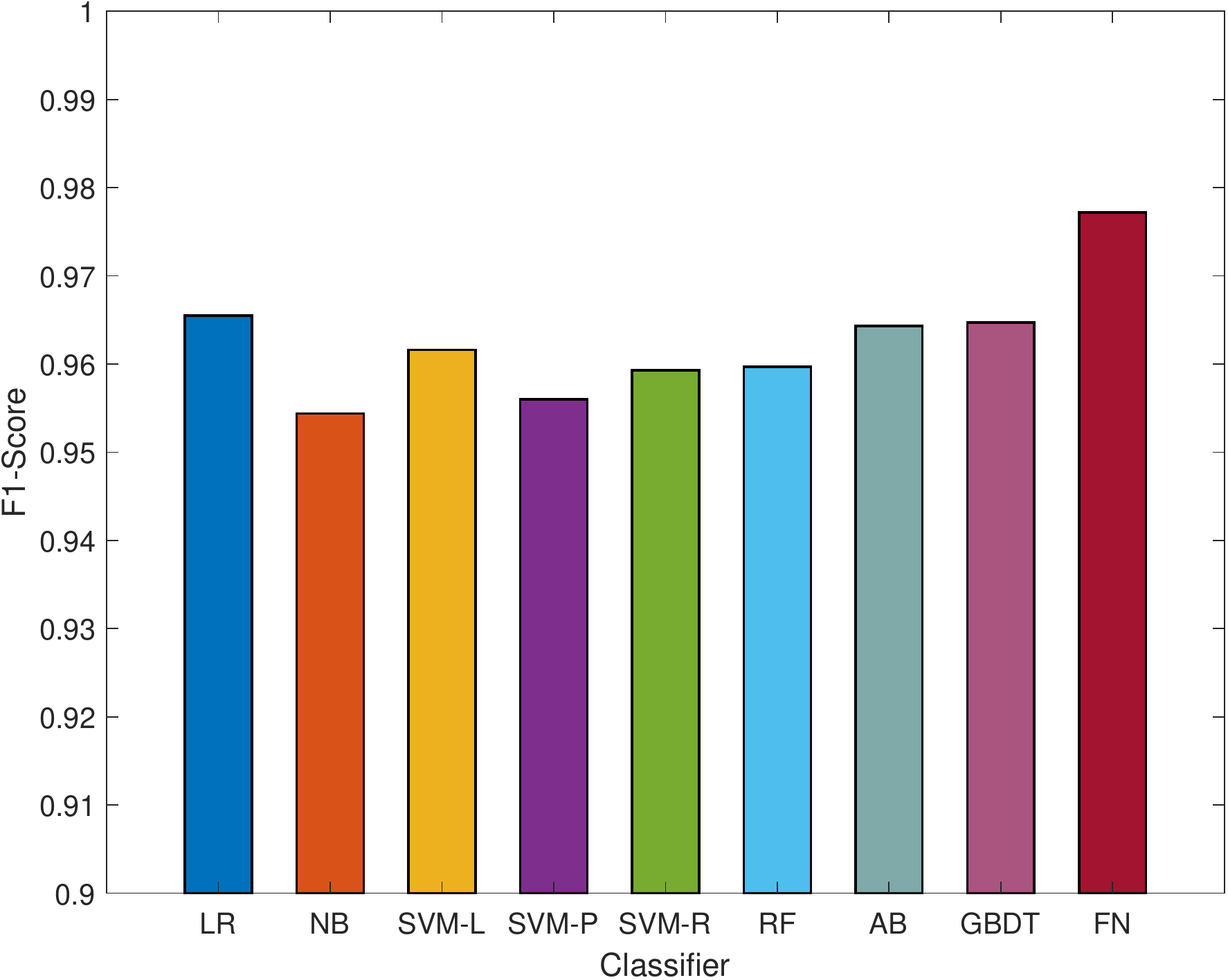}
    }
    \subfigure[Precision]{
        \includegraphics[scale=0.4]{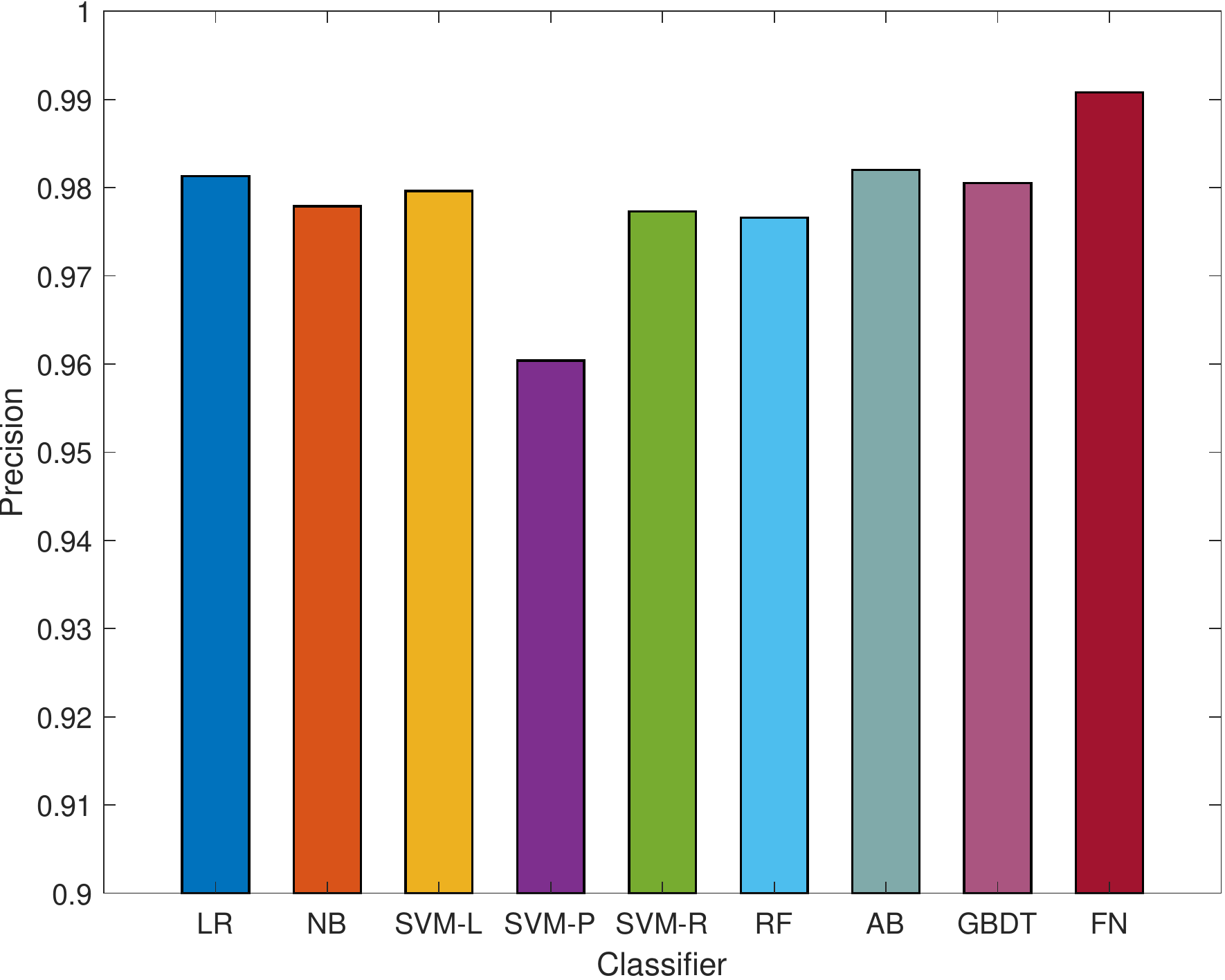}
    }
    \subfigure[Recall]{
        \includegraphics[scale=0.4]{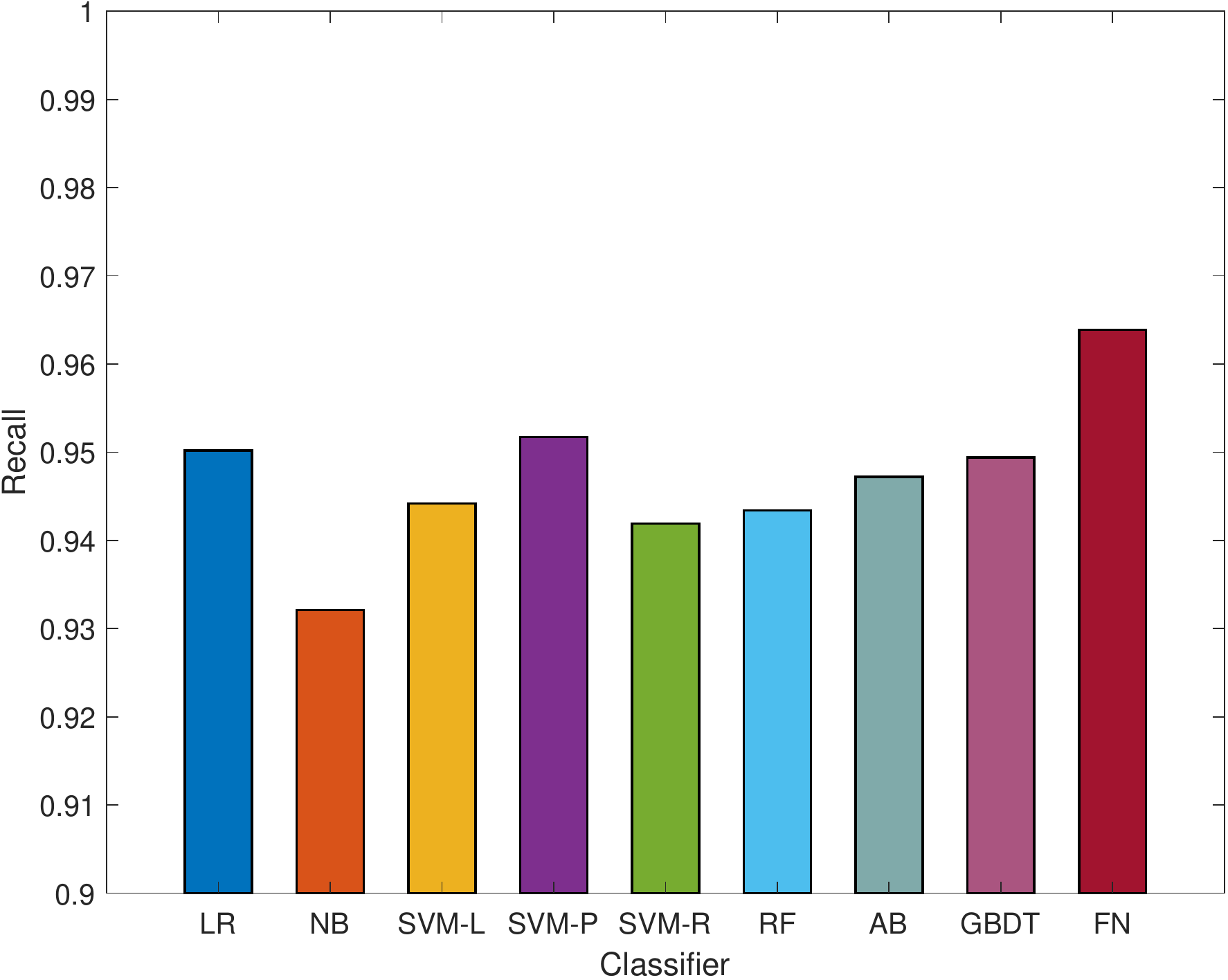}
    }

    \caption{Performance of Target Classifiers}
    \label{target-classifier}
\end{figure*}

Thus, multimodel classifiers are constructed by both the word vector classification network and the baseline statistical feature classifier. Table \ref{classifier-target-metrics} gives detailed metrics of all the target classifiers, while Fig. \ref{target-classifier} visualizes these metrics.

\begin{table}[htbp]
    \centering
    \caption{Classification Performance of Target Classifiers}
    \begin{tabular}{ccccc}
        \hline
        \textbf{Classifier}                            & \textbf{Accuracy} & \textbf{F1-Score} & \textbf{Precision} & \textbf{Recall} \\ \hline
        LR            & 0.9660 & 0.9655 & 0.9813 & 0.9502 \\
        NB & 0.9555 & 0.9544 & 0.9779 & 0.9321 \\
        SVM-linear & 0.9623            & 0.9616            & 0.9796             & 0.9442          \\
        SVM-poly                & 0.9562 & 0.9560 & 0.9604 & 0.9517 \\
        SVM-rbf              & 0.9600 & 0.9593 & 0.9773 & 0.9419 \\
        RF      & 0.9604 & 0.9597 & 0.9766 & 0.9434 \\
        AB              & 0.9649 & 0.9643 & 0.9820 & 0.9472 \\
        GBDT                     & 0.9653 & 0.9647 & 0.9805 & 0.9494 \\
        \textbf{FN (Proposed)}                         & \textbf{0.9775}   & \textbf{0.9772}   & \textbf{0.9908}    & \textbf{0.9639} \\ \hline
    \end{tabular}
    \label{classifier-target-metrics}
\end{table}

In the performance experiment of the target classifiers, each classifier reaches an F1-Score above 0.95. Particularly, our proposed FusionNet achieves the highest F1-Score value of 0.9772. It also obtains the highest value under all the other metrics. Compared to the second-highest LR, FusionNet improves its F1-Score by 1.21\%. Using transfer learning, the word feature is extracted as an input to the baseline statistical feature classifiers. However, transferring features through different classifiers may lead to a loss of information. Since multitask learning enables different tasks to share the same network structure and weights, it significantly reduces the loss of information caused by transfer learning, thus has better performance.

\begin{figure}[htbp]
    \centering
    \includegraphics[scale=0.5]{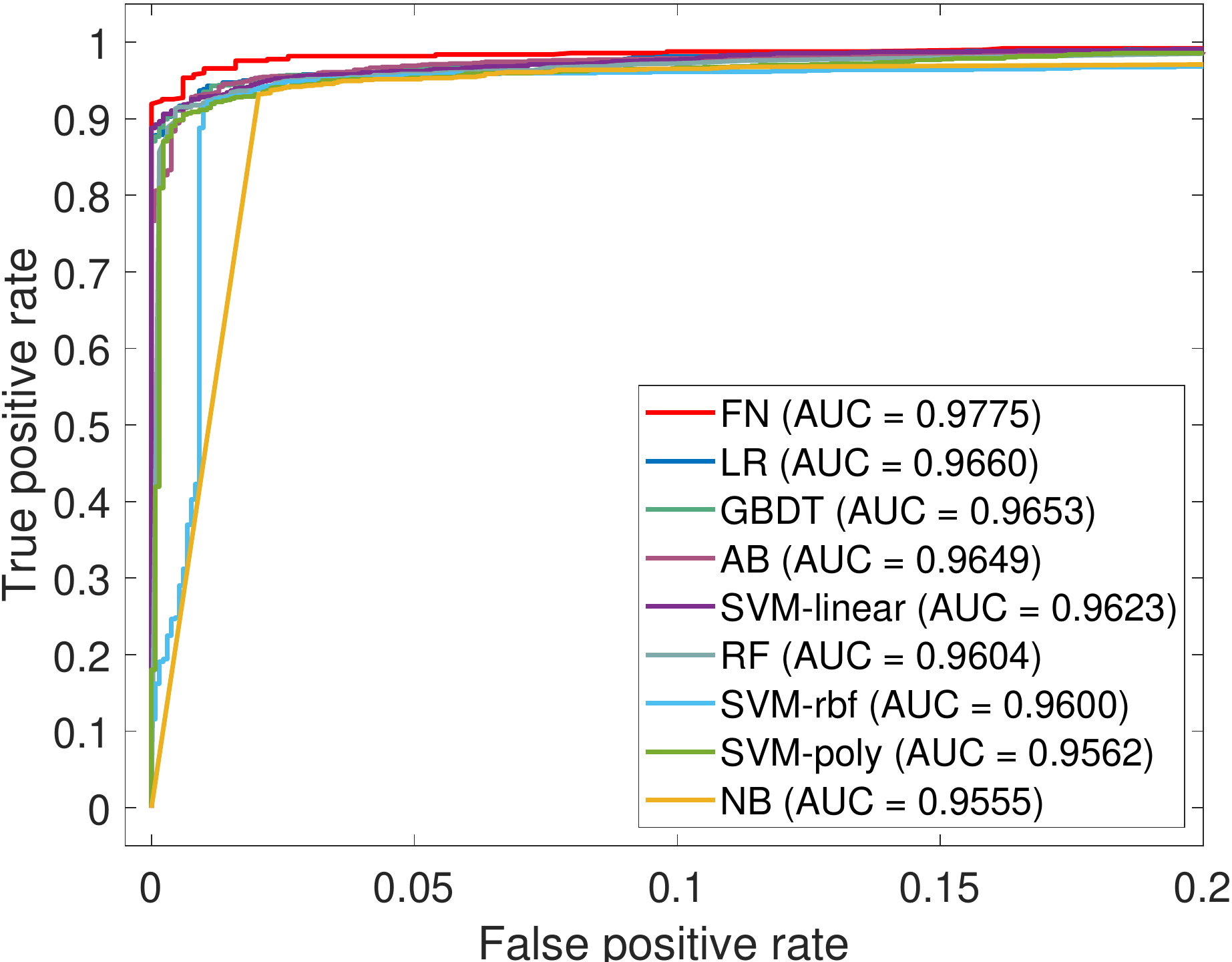}
    \caption{ROC Curves of the Target Classifiers}
    \label{target-roc}
\end{figure}

Furthermore, the ROC curves are given in Fig. \ref{target-roc}. We obtain the classification probability values of the model output and plot these ROC curves by sampling the multipoint FPR and TPR.

The result of the plot shows that all the curves are close to the upper left corner, which proves that all of the classifiers have excellent performance. The curve of FusionNet is closest to the upper left corner, achieving the best classification performance.

\subsection{Robustness of Unbalanced Training Samples}
In previous experiments, we used both depressed users and normal users with a proportion of 50\% in dataset $D_1$, $D_2$, and $D_3$. However, in the real OSN environment, depressed users exist in a minority of the whole user community. Due to the difficulty of collecting depressed user data, it is hard to ensure that the training and optimization process of the classifier can fully guarantee a balanced data proportion.

Therefore, by changing the proportion of depressed user samples (denoted as $\rho$), we analyze the F1-Score fluctuations on the target classifiers to evaluate its robustness of training an unbalanced number of samples. Each classifier is treated as a group. For each group, we will test nine values of $\rho$ from 0.1 to 0.9, with an interval of 0.1.

Here, we implement a new metric, namely the Intra-group F1-Score Variance (IFV), to calculate the variance of the F1-Score in each group. First, for each group, the mean value of the F1-Score is calculated and represented by $\overline{X}_{IF}$. The number of $\rho$ values taken in each group is noted as $T$. Therefore, the IFV metric is defined as:
\begin{equation}
   IFV = \sqrt{\frac{1}{T} \times \sum_{i=1}^{T}({F1}_i-\overline{X}_{IF})^{2}}
\end{equation}

\begin{table}[htbp]
    \centering
    \caption{IFV of Target Classifiers}
    \begin{tabular}{cc}
        \hline
        \multicolumn{1}{c}{\textbf{Classifier}} & \textbf{Intra-group F1-Score Variance (IFV)} \\ \hline
        LR                                      & 3.30e-5                          \\
        NB                                      & 1.82e-5                          \\
        SVM-linear                              & 4.60e-5                          \\
        SVM-poly                                 & 5.89e-5                          \\
        SVM-rbf                                 & 2.59e-5                          \\
        RF                                      & 2.80e-5                          \\
        AB                                      & 1.80e-5                          \\
        GBDT                                    & 2.92e-5                          \\
        \textbf{FN (Proposed)}                  & \textbf{1.01e-5}                 \\ \hline
    \end{tabular}
    \label{intragroup}
\end{table}

\begin{figure}[htbp]
    \centering
    \subfigure[F1-Score under Different $\rho$ Values]{
        \includegraphics[scale=0.4]{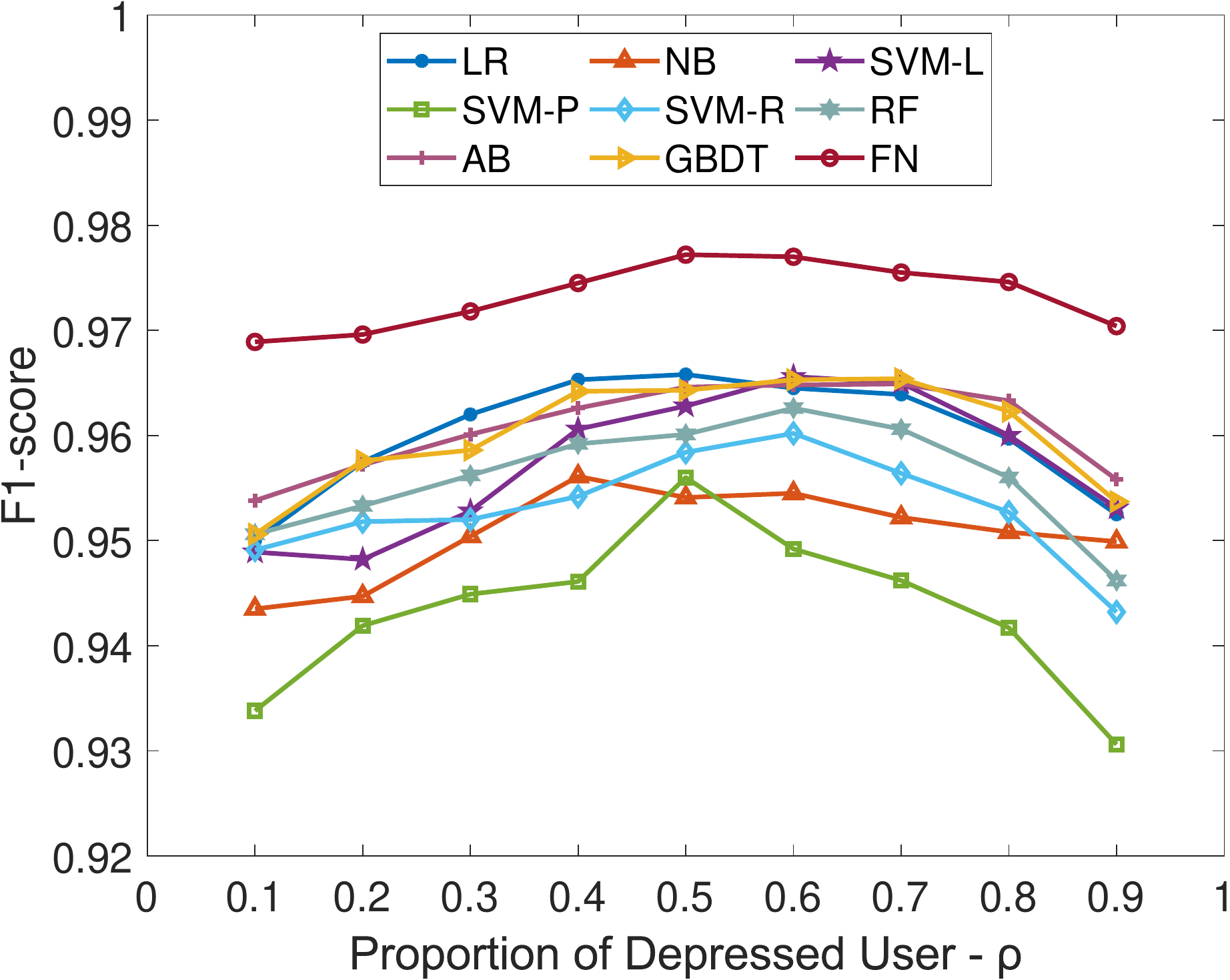}
    }
    \subfigure[IFV of the Target Classifiers]{
        \includegraphics[scale=0.4]{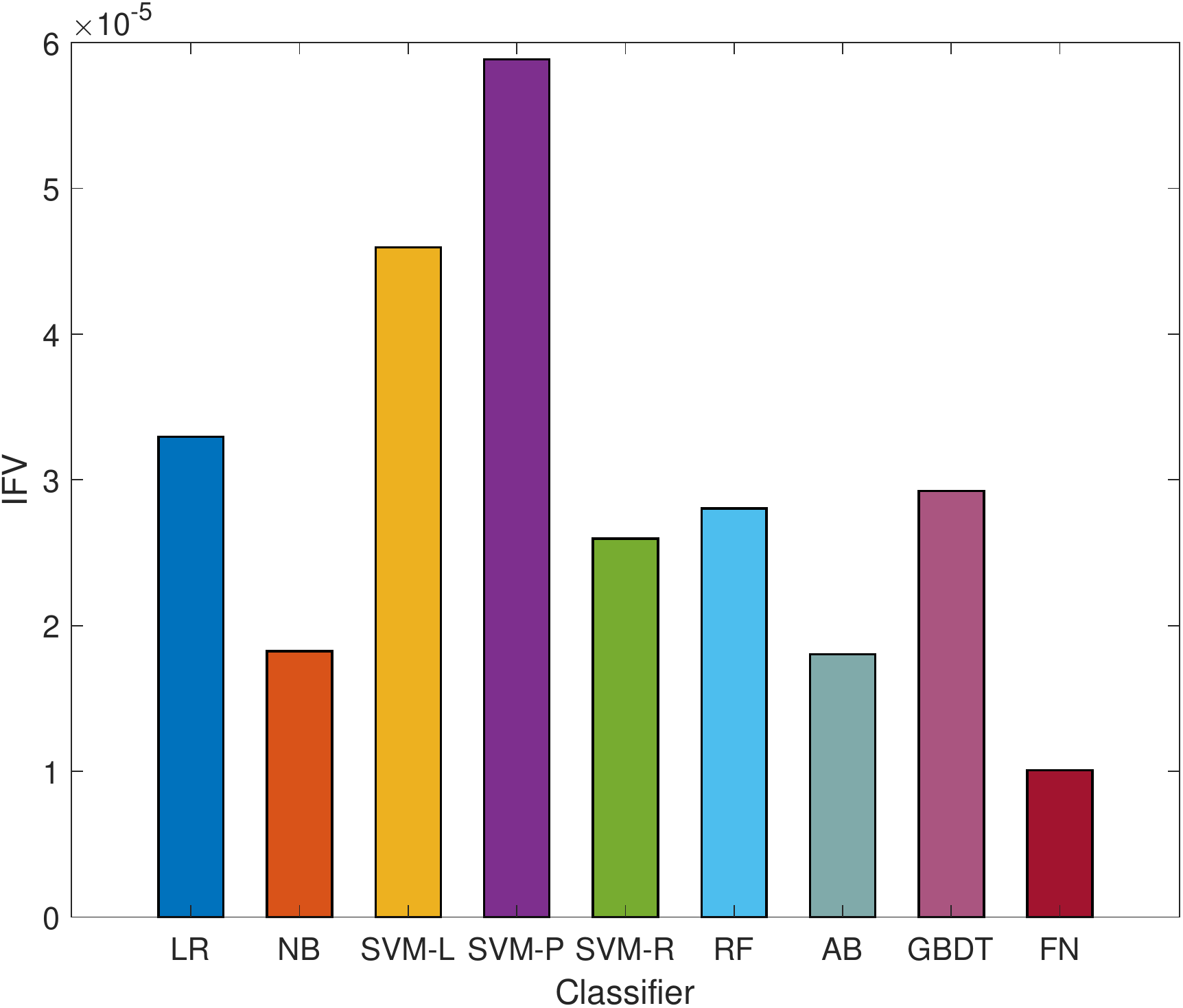}
    }
    \caption{Unbalanced Training Samples}
    \label{f-rho}
\end{figure}

Fig. \ref{f-rho}(a) shows the F1-Score of each classifier under different $\rho$ values. The experimental results show that when the proportion of data samples of depressed users and normal users are close to balance, the classifiers tend to achieve higher F1-Score.

Table \ref{intragroup} and Fig. \ref{f-rho}(b) shows the IFV metric of the target classifiers. Although LR once achieved a high F1-Score in the experiment of target classifiers, it has relatively low robustness for the unbalanced data due to the relatively poor decision ability of the single classifier. With the kernel learning strategy, the two types of SVM classifiers have a better ability to fit the data, but the values of F1-Score also fluctuates obviously when the training samples are unbalanced. Ensemble classifiers including RF and GBDT obtains better robustness in the experiment. The IFV metric of our proposed FusionNet reaches the minimum value among the target classifiers, indicating that FusionNet has the best robustness, i.e. the most stable classification performance.

In addition to the advantages of the multitask learning strategy mentioned in the previous part, we believe that the adaptive learning rate of Nadam can also help the FusionNet find the global optimal solution more quickly even if two classes of training data are not balanced.

\section{Conclusion}
In this work, we proposed a multitask learning-based approach to predict depressed users on Sina Weibo.

First, based on data collection and script filtering and manual labeling, we built and publish a large Weibo User depression detection dataset - WU3D. The total number of user samples reaches over 30,000 and each user has enriched information fields. This dataset will be quite sufficient to be used by subsequent researchers to complete further research.

Secondly, we summarized and manually extracted ten statistical features including text, social behavior, and picture-based features. The experimental results showed that all of them have varying degrees of distribution differences between normal users and depressed users, which can contribute positively to classification tasks. Our experimental results also proved that the feature engineering process of text information is the most vital part of depression detection on OSN.

Furthermore, we evaluated the performance of the pretrained model XLNet as the embedding model to solve downstream classification tasks. It showed that when the appropriate embedding length is selected, XLNet has excellent performance and efficiency in handling long text sequences.

Finally, we implemented a multitask learning DNN classifier, FusionNet, to simultaneously handle the word vector classification task and the statistical feature classification task. Benefit from the strategic advantages of multitask learning, FusionNet reduced the loss of feature information caused by transfer learning. Compared with the commonly used models in existing work, FusionNet has achieved a very significant performance improvement with an F1-Score of 0.9772 and showed the best classification robustness when the training samples are unbalanced. Thus, it has proven to be an ideal classification model when dealing with multiple classification tasks at the same time.

For future work, two directions will be further explored.
\textbf{(i)} The size of the dataset will be further expanded. Larger datasets will be constructed for training and evaluating classifiers to achieve better generalization performance. \textbf{(ii)} The characteristics and behavior patterns of depressed users will be further analyzed. We will propose more effective feature solutions for user-level depression detection on the OSN.

\bibliographystyle{IEEEtran}
\bibliography{depression}

\end{document}